\PassOptionsToPackage{table}{xcolor}
\documentclass[sigconf]{acmart}

\AtBeginDocument{%
  \providecommand\BibTeX{{%
    Bib\TeX}}}

\copyrightyear{2026}
\acmYear{2026}
\setcopyright{cc}
\setcctype{by}
\acmConference[CHI '26]{Proceedings of the 2026 CHI Conference on Human Factors in Computing Systems}{April 13--17, 2026}{Barcelona, Spain}
\acmBooktitle{Proceedings of the 2026 CHI Conference on Human Factors in Computing Systems (CHI '26), April 13--17, 2026, Barcelona, Spain}
\acmDOI{10.1145/3772318.3791585}
\acmISBN{979-8-4007-2278-3/2026/04}

\hypersetup{
    colorlinks=true,
    linkcolor=blue,
    filecolor=magenta,      
    urlcolor=cyan,
    pdftitle={Paper Critique},
    pdfpagemode=FullScreen,
    }
\usepackage[utf8]{inputenc} % For input encoding
\usepackage[T1]{fontenc}    % For font encoding
\usepackage{amsmath}        % For math environments
\usepackage{graphicx}       % For including images
\usepackage{enumitem}       % For custom lists
\usepackage{hyperref}       % For hyperlinks
\usepackage{natbib}         % For citation management
\usepackage{multirow}
\usepackage{rotating}
\usepackage{array}
\usepackage{booktabs}
\usepackage{amsthm}
\usepackage{float}
\usepackage{acmart-taps}
\usepackage{makecell}
\usepackage{geometry}
\usepackage{colortbl}
\usepackage{float}
\begin{document}

\title[Relational Sovereignty --- A New Telos for Socially Assistive Technology]{From Autonomy to Sovereignty ~---~ A~New~Telos~for~Socially~Assistive~Technology}

% title about operationalizing interdependence?

\author{JiWoong (Joon) Jang}
\email{jwjang@cmu.edu}
\orcid{0000-0003-0469-9501}
\affiliation{%
  \department{Human Computer Interaction Institute}
  \institution{Carnegie Mellon University}
  \city{Pittsburgh}
  \state{PA}
  \country{USA}
}

\author{Patrick Carrington}
\email{pcarrington@cmu.edu}
\orcid{0000-0001-8923-0803}
\affiliation{%
  \department{Human Computer Interaction Institute}
  \institution{Carnegie Mellon University}
  \city{Pittsburgh}
  \state{PA}
  \country{USA}
}

\author{Andrew Begel}
\email{abegel@cmu.edu}
\orcid{0000-0002-7425-4818}
\affiliation{%
  \department{Software Social Systems Department}
  \institution{Carnegie Mellon University}
  \city{Pittsburgh}
  \state{PA}
  \country{USA}
}

\renewcommand{\shortauthors}{Jang et al.}
\newcommand{\iref}[1]{~(§\ref{#1})}
\newcommand{\ilink}[1]{~(§\ref{#1})}
\theoremstyle{definition}
\newtheorem{defn}{Definition}
\begin{abstract}
Social accessibility research faces a persistent tension: assistive technologies (AT) predominantly pursue independence, yet disabled people's experiences reveal rich preferences for interdependence. 
Our analysis of 90 papers from 2011-2025 uncovered that this stems from a deeper issue---which crystallized through dialogue with three bodies of theories: (1) self-determination theory (SDT), (2) symbolic interactionism, and (3) posthumanist perspectives and crip technoscience. 
SDT illuminates individual needs; symbolic interactionism addresses construction of social meaning and stigma; Posthumanist and crip technoscience together challenges normalcy, governance, and the human-machine boundary. 
Through their tensions, we identify \textbf{\textit{relational sovereignty}} as an alternative telos---or goal---to autonomy. While our corpus equates autonomy with independence, sovereignty centers the power to choose between independence and interdependence. To operationalize this shift---from ``Can they do it?'' to ``Do they get to decide?''--- we introduce the Relational Sovereignty Matrix and four design interventions: (1) a sovereignty-centered reframing of SDT, (2) generative questions for justice-oriented reflection, (3) the idea of building through sovereign technical primitives, and (4) explicit consideration of power in AT design.
\enlargethispage{\baselineskip}
\end{abstract}

%   \includegraphics[width=\textwidth]{images/hero-image.png}
%   \Description{A hand drawn picture of crystals, some alone, some together.}


\maketitle
\section{Introduction}
\label{sec:introduction}
% introduction of the functional telos
Assistive Technology (AT) development has long pursued functional independence as its primary goal. The 1988 Technology-Related Assistance Act defined AT as tools to ``increase, maintain, or improve functional capabilities''~\cite{Harkin1988-ny, Zallio2022-xz}, a framing expanded by Daniels' healthcare justice theory which prioritizes ``normal species functioning''~\cite{Daniels2001-ah, Silvers2022-cm}. This paradigm operationalizes success as independence—completing tasks without relying on others~\cite{Shinohara2012-gd, Shinohara2018-eo}.

% Assistive Technology (AT) development has long been guided by a functional telos (goal) of independence. One of the earlier instances of the term includes the Technology-Related Assistance for Individuals With Disabilities Act, passed in 1988 in the United States, which identifies ATs as ``any item, piece of equipment, or product system [\ldots] that is used to increase, maintain, or improve functional capabilities'' of individuals with disabilities~\cite{Harkin1988-ny, Zallio2022-xz}. This teleological commitment was expanded in Norman Daniels' theory of healthcare justice, which establishes a moral priority to curing disease and disability, framing ATs as necessary to effectuate ``normal species functioning'' (though still secondary to removing the disability entirely)~\cite{Daniels2001-ah, Silvers2022-cm}. This principle of AT design in HCI has been operationalized for years as the pursuit of independence, often posed as reducing or eliminating a disabled person's reliance on others, framing self-sufficiency as the ultimate success~\cite{Shinohara2012-gd, Shinohara2018-eo}. 

% other telos from DS, DJ, Critical studies
While the functional telos (or ultimate goal) has structured policy, research, and evaluation, HCI has absorbed countervailing commitments from disability studies, critical disability studies, and disability justice that reorient ends toward remaking relations, meanings, and power~\cite{Mankoff2010-lb,Williams2019-bp,Spiel2020-ns,Williams2021-se,Sum2022-lp,Hofmann2020-my}. These perspectives advance the social model of disability~\cite{Mankoff2010-lb,Williams2021-se,Hofmann2020-my}, foreground stigma and identity as design materials~\cite{Garland-Thomson2009-pa,Williams2019-bp}, advocate for participatory inclusion~\cite{Mankoff2010-lb,Spiel2020-ns,Ymous2020-gu,Hofmann2020-my}, and center interdependence, collective access, and care as positive, chosen arrangements~\cite{Invalid2017-zi,Mingus2017-uu,Bennett2018-pz,Sum2022-lp}. Altogether, these commitments replace a single, normative end with a spectrum---from individual self‑direction to collective, chosen interdependence---turning the placement of AT along this spectrum into an explicit consequential design choice.
% \enlargethispage{\baselineskip}

% what does this work seek to do? present the RQ
% CONDENSED VERSION:
The work for this paper began with an unrelated exploratory question: What constitutes the work of social accessibility? Through systematic analysis of 90 papers, we discovered a persistent pattern: \textit{disabled people faced unresolved tensions between independence and interdependence} that existing frameworks could not adequately address. This led us to ask: \textbf{How should Assistive Technology be positioned along the independence-interdependence spectrum?}

The field lacks a unifying framework. Bennett et al.~\cite{Bennett2018-pz} articulated interdependence as complementary to independence, not a replacement, while documenting both empowering and paternalistic forms in practice. Recent work reveals that pursuing full independence can strip users of control, while uncritical interdependence risks paternalism~\cite{Xiao2024-fe, Bhattacharjee2019-sk}. The consequences of this ambiguity are not mere theoretical: over‑prioritizing independence can privatize access labor, while uncritical interdependence can lapse into coerced reliance or dependence. This reveals the need for principled ways to reason about \textit{when}, \textit{how}, and \textit{on whose terms} to support each mode.

To pursue this question, we focus on \textit{social accessibility}: a subset of HCI where the independence-interdependence dynamic is made visible in practice. By social accessibility, we mean research which study technologies and practices to support disabled people's social participation: how devices are perceived by others, how they mediate impression management and disclosure, how they afford identity expression and aesthetics, and how they redistribute the social labor of access. Rather than treating ``the social'' as a side effect of function, social accessibility foregrounds these dynamics as primary design materials and outcomes~\cite{Shinohara2017-md, Shinohara2018-gu, Shinohara2018-eo, Shinohara2012-gd}. Social accessibility by construction is therefore a site where decisions about both independence and interdependence are actualized and experienced---serving as an appropriate case study for our needs~\cite{Shinohara2011-bf, Shinohara2018-gu}. 

% what we did

To explore how social accessibility is conceptualized and enacted in HCI, we conducted a constructive grounded analysis~\cite{Charmaz2017-av, Charmaz2017-cs, Charmaz2006-ty} of 90 theoretically rich papers (2011–2025) curated from a systematically derived corpus of 605 social accessibility relevant publications across premier HCI venues~(§~\ref{subsec:corpus-construction}). Initially, rather than testing hypotheses, we sought to understand what patterns would emerge from examining how the field approaches the social dimensions of assistive technology. In our selection of these works, we prioritized works that provided situated accounts of use and design (e.g., qualitative studies, design cases, theory‑building) and excluded purely algorithmic or instrumentation papers. 
% say what the analysis *did do*
Our inductive synthesis surfaced recurrent tensions that the independence-interdependence spectrum alone could not resolve: autonomy was often invoked to mean independence rather than volition---the ability to act according to one's own values and interests; interdependence appeared as both empowering support and coerced reliance. We found many accounts that hinted at asymmetries of power, raising questions of who defines goals, whose terms prevail, even when collaboration was present~\cite{Xiao2024-fe}.

To interpret these tensions, we then performed a meta-synthesis, placing our empirical findings in productive and diffractive~\cite{Barad2009-qs} dialogue with three theoretical lenses (which were themselves in tension with one another). With these lenses, we ask:

\begin{itemize}
    \item \textbf{The Self}: What does volition require, and when is it supported or undermined?
    \item \textbf{The Interaction}: How are meanings of disability and ability negotiated in situated encounters?
    \item \textbf{The Body}: How are bodies configured by technology, and who governs the terms?
\end{itemize}

We drew on self-determination theory (SDT) to address the first, symbolic interactionism (SI) to address the second, and perspectives from posthumanism and crip technoscience to address the third. From this synthesis, we propose \textit{relational sovereignty}---the recognized power to choose one's relational mode and to define the telos of social engagement---as a more precise alternative to traditional autonomy.

% what is sovereignty?
\subsection*{Sovereignty: Not Just Which Relational Mode, But on \textit{Whose} Terms?}
% MUCH SHORTER VERSION:
Relational sovereignty refers to the disabled person's authority to decide when to seek independence or interdependence (relational mode), and on what terms. We adopt this term with care. \textit{Sovereignty} originates from Indigenous sovereignty movements, which assert self-determination over governance~\cite{Barker2005-xs, Alfred2005-li}, and data sovereignty frameworks like OCAP (Ownership, Control, Access, Possession)~\cite{FNIGC-sp} and CARE principles~\cite{Carroll2020-mq}. 

We invoke sovereignty relationally and at the interpersonal / infrastructural scale to name who sets aims and boundaries in AT use, not to displace those political claims. Within disability justice, this meaning is already implicit: ``Nothing about us without us'' asserts definitional authority~\cite{Spiel2020-ns}; collective access and access intimacy name chosen, directed interdependence~\cite{Invalid2017-zi,Mingus2017-uu}. Our usage of sovereignty focuses these threads into a single criterion: whether the disabled user’s aims and relational terms are recognized and upheld.

We operationalize this concept through the Relational Sovereignty Matrix~(Figure~\ref{fig:relational-sov-matrix}), an analytical tool that classifies user experiences along the orthogonal axes of Relational Mode (from Independence to Interdependence) and Sovereignty (from Conditional to Recognized). In this framing, dependence is not a relational mode but a sovereignty condition: it denotes externally imposed terms---whether one is ``alone'' with a system or ``assisted'' by people. The remainder of this paper is dedicated to building and exploring the relational sovereignty framework.
% We document teleogical history behind AT~(§~\ref{sec:related-works}), and detail the methodology used to derive it~(§~\ref{sec:methods}). Then, in we document our findings from our analysis of social accessibility works and motivate the need for sovereignty~(§~\ref{sec:findings}), We formally introduce the theory of sovereignty~(§~\ref{sec:theory-of-sovereignty}), unpack the matrix, and use rich examples from our corpus to illustrate its four resulting quadrants. Finally, we discuss the implications of this model~(§~\ref{implications}) and propose a set of interventions for designing sovereign technologies.
This paper traces AT's teleological evolution~(§~\ref{sec:related-works}), details our methodology~(§~\ref{sec:methods}), presents findings motivating sovereignty~(§~\ref{sec:findings}), introduces the theoretical framework~(§~\ref{sec:theory-of-sovereignty}), and proposes implications and design interventions for designing sovereign technologies~(§~\ref{sec:implcations}).

\section{Background and Related Works}
\label{sec:related-works}
The field of accessibility and assistive technology has undergone profound shifts in how it thinks about its goals and purposes. What began as a primarily functional enterprise has evolved into a complex negotiation of individual needs, social meanings, and values. This evolution, however, has created new challenges, including how to honor both individual autonomy and collective interdependence without compromising either. This section traces these intellectual developments to establish why new theoretical developments are needed.

\subsection{The Evolving Telos of Assistive Technology}
The goals driving assistive technology development have transformed dramatically over the past several decades, reflecting broader shifts in how society conceptualizes disability itself.

\subsubsection{The Functional Independence Paradigm (Pre-2010s)}

Historically shaped by rehabilitation engineering and medical models of disability, assistive technology emerged with a singular focus on functional restoration. The 1988 Technology-Related Assistance Act codified this view, defining AT as tools to ``increase, maintain, or improve functional capabilities''~\cite{Harkin1988-ny}. This framing positioned disability as individual deficit requiring technological remedy, with success measured by approximation to normative functioning~\cite{Cook2007-cx, Daniels2001-ah}. 

Within this paradigm, independence became the ultimate goal: not merely the ability to complete tasks, but to do so without relying on others. Early HCI accessibility research inherited these assumptions, prioritizing technical solutions that enabled disabled users to operate autonomously within existing systems. The field's metrics reflected these values: task completion rates, error reduction, and time-to-independence dominated evaluation frameworks.

\subsubsection{The Social Turn (2010–2015)}
The early 2010s marked a shift as HCI began absorbing insights from disability studies. Pullin's \textit{Design Meets Disability}~\cite{Pullin2021-xo} challenged the field to consider aesthetics and identity expression as central, not peripheral, to AT design. Mankoff, Hayes, and Kasnitz~\cite{Mankoff2010-lb} advocated for importing disability studies perspectives into HCI, to interrogate the medical model's assumptions and center disabled people's own research agendas.

This time also saw the emergence of \textit{social accessibility} as a distinct research program. Shinohara and Wobbrock's foundational work~\cite{Shinohara2011-bf, Shinohara2017-md} demonstrated that assistive devices carry social meanings beyond their functional utility—they mark users as different, require complex impression management, and can simultaneously enable and stigmatize. Their research revealed that technically effective solutions could still fail socially, creating new forms of exclusion even while addressing functional needs.

Parallel scholarship began documenting the hidden social infrastructures underlying ``independent'' technology use. Branham and Kane~\cite{Branham2015-rr} exposed the invisible labor that friends and family perform to maintain accessible systems. The DIY and maker movements in accessibility, documented by Hurst and Kane~\cite{Hurst2013-gn} and Hofmann et al.~\cite{Hofmann2016-wl}, revealed users as active creators rather than passive recipients, while also showing how ``independent'' making relied on community knowledge, shared designs, and collective problem-solving.
\enlargethispage{1.5\baselineskip}
\subsubsection{Toward Justice and Liberation (2015–Present)}
Recent scholarship has pushed beyond inclusion toward more transformative goals, influenced by the Disability Justice movement and critical theories. This shift recognizes that making technology accessible within unjust systems may perpetuate rather than challenge oppression. Intersectional analyses revealed how accessibility efforts can reinforce other forms of marginalization. Williams et al.~\cite{Williams2019-bp} documented how assistive technologies became tools of racialized surveillance. Costanza-Chock's \textit{Design Justice}~\cite{Costanza-Chock2018-jg} framework showed how even participatory design could exclude multiply-marginalized disabled people when power dynamics go unexamined.

Building on these critiques, scholars from disability studies and STS have articulated more expansive visions. Hamraie's \textit{Building Access}~\cite{Hamraie2017-cq} positions crip technoscience as a form of world-making that reimagines rather than accommodates. Hendren's \textit{What Can a Body Do?}~\cite{Hendren2020-co} inverts the traditional AT paradigm, asking how environments might adapt to bodies rather than demanding bodily adaptation. Williams et al.~\cite{Williams2021-se} synthesizes these threads into a vision of crip HCI that embraces non-normative futures. Shew's \textit{Against Technoableism}~\cite{Shew2023-fm} fundamentally challenges the assumption that technology should ``solve'' disability at all, fully rejecting Daniels' healthcare justice paradigm~\cite{Daniels2001-ah}.

These perspectives raise teleological questions about AT: Should technology help users achieve independence within existing systems, or foster interdependence that challenges those systems? When is self-reliance empowering versus isolating? When does support become paternalism? The field's inability to answer these questions reveals itself most clearly in the ongoing debate between independence and interdependence as design goals.

\subsection{The Independence-Interdependence Complex}
\label{sec:ind-inter-paradox}
Independence has long served as a central goal in assistive technology research and design~\cite{Shinohara2012-gd, Shinohara2018-eo}. Yet critical disability scholars and activists have challenged this framing, arguing that what we call "independence" typically obscures reliance on infrastructure, design choices, and invisible labor rather than eliminating reliance altogether~\cite{Mingus2017-uu, Hamraie2017-cq}. From this perspective, a power wheelchair user's "independent" mobility depends on paved roads, electricity, and maintenance systems—interdependence with built environments rather than with people.
This tension between independence as an orienting goal and interdependence as lived reality has generated significant work in HCI and accessibility research. We review this literature to understand how the field has grappled with these competing frames, attending particularly to how researchers have conceptualized the relationship between relational modes and user agency.

\enlargethispage{\baselineskip}
\subsubsection{The Turn to Interdependence}
Bennett et al.'s 2018 \textit{Interdependence as a Frame for Assistive Technology Research and Design}~\cite{Bennett2018-pz} served as a turning point for accessibility research. They identified a fundamental disconnect: while AT research pursued independence as its primary goal, disability communities had long embraced interdependence as both more accurate and more desirable. Their framework operationalized interdependence through four principles: (1) recognizing relations between people, technologies, and environments; (2) revealing mutual giving and receiving; (3) highlighting disabled people's contributions; and (4) challenging ability-based hierarchies.

In introducing interdependence, Bennett et al. positioned it as ``one possible orientation'' that ``opens new design possibilities,'' leaving open the question as to how to decide between them.

\subsubsection{Empirical Observations of the Complex}
Empirical work demonstrate both the promise and peril of interdependence. Studies of collaborative accessibility show that the same support relationships can be simultaneously empowering and constraining.

Branham and Kane's~\cite{Branham2015-rr} ethnography of blind-sighted pairs found ``collaborative accessibility'' where partners co-create access, but also documented the ``preparation paradox'' (how collaborative preparation enables later independence) alongside ``intervention tensions'' (where needed support causes frustration). Studies involving people with intellectual disabilities and their personal assistants reveal similar complexity: interdependent arrangements can foster belonging but also create vulnerability to others' schedules, attitudes, and continued presence~\cite{Dowling2019-uo}.

Wheelchair and mobility research demonstrates wheelchair users constantly switching devices and strategies based on environmental and social contexts, including instances where ``the effort invested in using an assistive technology was not worth the gain [\ldots] having a carer perform the task instead was preferential'' revealing complex independence-interdependence negotiations~\cite{Dorrington2016-pv, Khalili2022-qa}.

\subsection{Theoretical Resources for Reframing AT Goals} To move beyond documenting tensions toward resolving them, we need analytical tools that operate at multiple scales---from individual psychology to social interaction to political transformation.

Existing frameworks in HCI, such as Schneider et al.~\cite{Schneider2018-us}, have characterized empowerment discourse as split between \textit{power-to} (enhancing individual capacity and doing) and \textit{power-over} (redistributing authority and resisting domination). However, in the context of AT, this distinction often collapses: a user's power-to act independently is frequently contingent upon who holds power-over the infrastructure and social terms.

We therefore seek a theoretical framework that does not treat capacity and governance as separate tracks, but as mutually constitutive. To do so, we draw on three \textit{bodies of theory} that allow us to inspect both sides of this equation: 
\begin{itemize} 
\item \textbf{The Self} (Self-Determination Theory), which clarifies the conditions for individual volition (or power-to); 
\item \textbf{The Interaction} (Symbolic Interactionism), which reveals how authority is negotiated interpersonally; 
\item \textbf{The Body} (Posthumanism \& Crip Technoscience), which examines how governance is embedded in material sociotechnical arrangements (or power-over). 
\end{itemize}

These clusters of related perspectives share orienting questions but may contain internal debates, competing strands, and productive disagreements.
\subsubsection{The Self}
Our first lens attends to the psychological dimensions of agency: what individuals need to thrive and act with volition. We draw primarily on self-determination theory (SDT), a macro-theory of human motivation and personality composed of several interrelated mini-theories, with the overall goal of understanding the conditions that support or undermine people's innate tendencies toward growth, well-being, and self-motivation. In this work we focus on two mini-theories that are most salient: Basic Psychological Needs Theory (BPNT) and Organismic Integration Theory (OIT)~\cite{Deci1985-so, Ryan2000-de}. 

BPNT frames autonomy, competence, and relatedness as basic psychological needs whose satisfaction or frustration is strongly linked with well‑being, while OIT emphasizes that thwarting any of the three basic needs can produce controlled forms of motivation and ill-being, even when behavior seems successful on the surface. This is crucial for AT, where externally defined ``independence'' can co‑occur with deep needs frustration. We use these components of SDT not to claim universal needs in an absolute sense, but to diagnose how HCI work selectively invokes autonomy while sidelining relatedness and user-defined competence. 

SDT has become one of the dominant frameworks through which HCI engages with questions of the self---motivation, volition, and psychological well-being---making it a productive site for examining the field's assumptions. Yet recent work by Tyack and Mekler on SDT in HCI games research highlights how easily SDT can become an ``unquestioned paradigm'' when its mini-theories and normative commitments are only shallowly engaged~\cite{Tyack2024-gj}. In their expanded review of 259 SDT-based games papers, they find that HCI scholarship overwhelmingly treats SDT as a vocabulary and a source of ready‑made scales (PENS, IMI), while rarely using its propositions to guide research questions, contest findings, or ``talk back'' to the theory. Mini‑theories such as BPNT, OIT, or Causality Orientations Theory (COT) are seldom named, and autonomy is frequently mischaracterized as control or choice rather than volition, with relatedness and need frustration largely ignored. 

Following their critique, we treat SDT as one influential framework among others, and make explicit which parts we draw on and where our findings challenge how SDT is typically mobilized in HCI. For AT analysis, questions raised by SDT provide vocabulary to distinguish between independence (doing things alone) and autonomy (choosing one's relational mode). This distinction proves crucial: a user might choose interdependence while maintaining autonomy, or experience independence without autonomy when forced into self-reliance.

\subsubsection{The Interaction}
% symbolic interactionism is a theory that explains how individuals construct social reality through shared symbols, meanings, and interpretations~\cite{Goffman1959-uf}. Applied to AT, this lens reveals how devices become symbols that users and others interpret, requiring constant negotiation of stigma~\cite{Goffman1986-kp}, identity, and social norms~\cite{Shinohara2011-bf}. This framework illuminates the ``social work'' that AT requires and performs, including things like impression management, disclosure decisions, and relationship mediation.
Our second lens attends to how meaning, identity, and social reality emerge through situated encounters. We draw primarily on symbolic interactionism (SI), a family of theories for understanding how selves, roles, and social realities are constructed through interaction. Our use of SI aligns primarily with the Chicago School (e.g. Mead; Blumer), which emphasizes that meanings are not inherent in objects but arise through ongoing interaction, and that people act toward things---including technologies---on the basis of the meanings those things have for them~\cite{Mead1967-vq, Blumer1992-oz}. We take up key concepts from this lineage, for example the definition of the situation, joint action, and the emergent, reflexive self, to show how disability arises through interaction rather than inhering in individuals or environments. We also draw heavily on Goffman’s account of the interaction order and dramaturgy, particularly his analyses of stigma symbols, impression management, and front- and back-stage behavior~\cite{Goffman1959-uf, Goffman1986-kp}, which help us examine how assistive devices can both mark and be strategically managed in social life.

At the same time, we acknowledge more structural strands of SI, such as the Iowa School, (associated with Kuhn and others), which foreground relatively stable self‑concepts and role identities, and later negotiated‑order perspectives~\cite{Strauss1978-dq} that connect institutional structures with their continual renegotiation in everyday practice. Rather than treating these as incompatible camps, we use negotiated order as a bridge: assistive technologies are understood both as structural anchors that position disabled people in particular roles (e.g., “dependent patient,” “competent worker”) and as interactional props that users mobilize to negotiate, resist, and reconfigure those roles. This SI framing provides the vocabulary we need to analyze how the papers in our corpus depict identity work, disclosure, stigma management, and relational expectations around assistive technology.

\subsubsection{The (Material) Body}
Our third lens attends to bodies as materially configured sites entangled with technology---how bodies are designed for, contested, and politically situated. We draw on perspectives that challenge fundamental categories---human/machine, normal/abnormal, able/disabled---that other frameworks take for granted~\cite{Haraway2013-vs, Kafer2013-sp}. Though the traditions we include here disagree with one another in important ways, together they reveal how ``universal'' needs (and technologies built on said assumptions) often encode particular (usually abled) ways of being as natural and desirable~\cite{Williams2019-bp}. These perspectives in this body of theory push us to ask not just whether technology works, but \textit{whose} vision of ``working'' prevails and what alternative futures might be possible.

\paragraph{Posthumanist Perspectives}
Posthumanism challenges the liberal humanist assumption of a bounded, self-sufficient individual by emphasizing distributed agency and human-nonhuman assemblages. Haraway’s \textit{Cyborg Manifesto} destabilizes the human-machine boundary and foregrounds hybrid embodiments as already ordinary rather than exceptional~\cite{Haraway2013-vs}. Barad’s agential realism similarly treats agency as an enactment emerging through ``intra-action'' among bodies, devices, and environments rather than an intrinsic property of discrete individuals~\cite{Barad2009-qs}. Braidotti’s posthuman subject~\cite{Braidotti2013-wo} and Bennett’s notion of ``vibrant matter''~\cite{Bennett2020-lj} extend this line of thought by insisting that the capacities of any actor are co-constituted with infrastructures, objects, and ecologies. From this perspective, true or absolute independence is ontologically untenable: all action is necessarily entangled action.

Science and Technology Studies (STS) of disability ground these posthuman insights in mundane, everyday care practices. Winance’s analysis of wheelchair use, for example, traces how bodies, devices, and built environments are mutually configured over time, revealing independence as an achievement of an entire assemblage coming together rather than an attribute of a lone user~\cite{Winance2006-sd}. Moser shows how disabled people and technologies co-produce one another in everyday life, complicating simple binaries of dependence versus autonomy~\cite{Moser2005-ni}. Mol and Pols likewise describe care as a distributed accomplishment in which patients, professionals, and artifacts continually re-negotiate what ``good'' care is and who does what work~\cite{Mol2002-th, Pols2012-ad}. Taken together, these posthuman and STS accounts remind us that what HCI often calls ``independence'' already rests on dense, largely invisible networks of human and nonhuman support.

\paragraph{Crip Technoscience}
Crip technoscience and critical disability studies begin from disabled people’s situated interdependence rather than from abstract human subjects. Hamraie and Fritsch’s \textit{Crip Technoscience Manifesto} articulates disabled makers and users as technoscientific agents whose practices expose and rework the ableist assumptions built into environments and devices~\cite{Hamraie2019-fx}. Mingus’ writing on access intimacy and interdependence names forms of care and relational access that are both materially specific and politically charged, rejecting the myth of independence while insisting on disabled people’s authority to direct the relations they rely on~\cite{Mingus2017-uu}. Shew’s critique of technoableism further challenges imaginaries in which technology is tasked with ``fixing'' disability or restoring users to normative functioning, arguing instead for centering disabled ways of being and resisting compulsory cure and enhancement~\cite{Shew2023-fm}.

\paragraph{Viewing Posthumanist and Crip Perspectives Diffractively}
These crip perspectives also interrogate some strands of posthuman and feminist theory. Kafer's work on crip futurity shows how cyborg and posthuman narratives often presume an able-bodied subject or treat disability as a deficit to be overcome, erasing disabled people's existing, often coercively governed, relationships with technology~\cite{Kafer2013-sp}.  Weise’s direct critique of Haraway's cyborg visions~\cite{Weise2018-ge} and Forlano’s autoethnography of chronic illness~\cite{Forlano2017-cv} similarly highlight how techno-utopian visions of a seamless human-machine integration can obscure asymmetrical power relations, surveillance, and labor involved in actually maintaining disabled life. Crip technoscience thus adds an explicitly political and normative layer to posthuman entanglement: it asks whose bodyminds are centered, which dependencies are valued or stigmatized, and who controls the terms of technological mediation~\cite{Price2015-ty}.

We recognize these fundamental differences and therefore read Posthumanism and Crip Technoscience \textit{diffractively rather than as a synthesis}~\cite{Geerts2016-wu, Barad2009-qs}. Posthuman and feminist new materialist accounts decenter the autonomous liberal subject and make visible the assemblages involved, while crip technoscience and critical disability theory insist that those assemblages are sites of technoableism, contestation, and world-building. STS studies of disability such as Winance, Moser, Mol, and Pols provide connective tissue between these traditions by showing how bodies, devices, and institutions are co-shaped in practice. Where posthumanism suggests that we are always already entangled, crip technoscience asks who governs those entanglements, to what ends, and with what recourse. It is this governance question that we develop with the term \textit{relational sovereignty}.

\subsubsection{Productive Tensions Between Lenses}
\label{sec:productive-tensions}
We utilize these three bodies of theory as analytic lenses because their worldviews conflict, generating insight through their incompatibilities. First, SDT posits universal psychological needs, yet crip theory interrogates how these ``universal'' standards of competence often encode ableist norms as natural~\cite{Deci1985-so, Shew2023-fm}. Second, while SDT centers the bounded agent’s volition, posthumanist perspectives reframes the individual into distributed assemblages, raising critical questions about who governs the ``independence'' provided by infrastructure~\cite{Barad2009-qs, Mingus2017-uu}. Third, symbolic interactionism reveals how users negotiate meanings and manage audiences to mitigate stigma, yet Crip Technoscience highlights that social negotiation often faces a wall of material constraints---such as platform defaults or algorithmic biases---that are not negotiable practically~\cite{Blumer1992-oz, Hamraie2019-fx}.

These conflicts extend to the temporal and relational dimensions of access. The interactional pressure to maintain normative ``flow'' (SI)~\cite{Goffman1959-uf} explicitly conflicts with the need for ``crip time'' (crip theory) and rest~\cite{Kafer2013-sp}, just as the impulse to manage others as an audience (SI) is in tension with the need to embrace them as allies in collective access (Disability Justice)~\cite{Invalid2017-zi}. \textit{Relational sovereignty} emerges from these frictions as the necessary power to arbitrate them: the recognized authority to set the terms of one's own pacing, definitions, and relationships against the inertia of normative systems.

\begin{figure*}[t!]
    \centering
    \includegraphics[width=\textwidth]{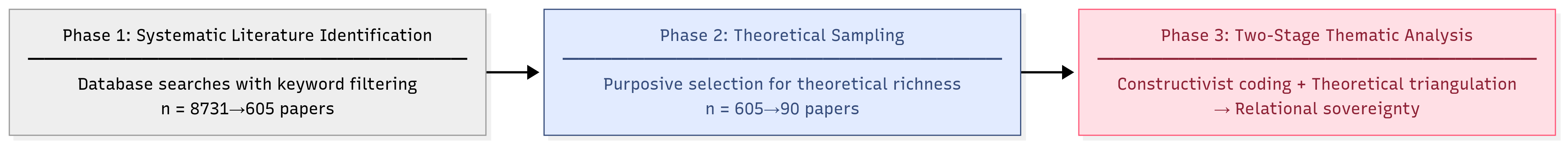}
    \caption{Stages of the Methodological Process: From keyword searches, corpus curation, and multi-stage synthesis.}
    \label{fig:method-phrases-diagram}
    \Description{Flow diagram showing three sequential phases of the research methodology, connected by arrows from left to right. Phase 1: Systematic Literature Identification, involving database searches with keyword filtering, resulting in n=605 papers. Phase 2: Theoretical Sampling, using purposive selection for theoretical richness, narrowing to n=90 papers. Phase 3: Two-Stage Thematic Analysis, combining constructivist coding with theoretical triangulation, leading to the sovereignty framework. Each phase is represented as a rectangular box with the phases connected by directional arrows indicating the sequential progression of the methodology.}
\end{figure*}

\subsection{Indigenous Scholarship on Sovereignty, Disability, and Data Governance}
Indigenous sovereignty represents the inherent right of Indigenous peoples to self-determination and self-governance over their territories, resources, and ways of life~\cite{Barker2005-xs, Alfred2005-li}. This concept emerged from resistance to colonial governance structures that sought to eliminate Indigenous political authority and impose external control~\cite{Barker2005-xs}. Data sovereignty extends sovereignty principles to the digital realm, asserting that data about Indigenous peoples and territories should be subject to Indigenous governance and laws~\cite{Kukutai2016-nf}. This movement emerged in response to centuries of extractive research where Indigenous knowledge, genetic materials, and cultural information were taken without consent, often used to the detriment of Indigenous communities~\cite{Alfred2005-li, Kukutai2016-nf}.

Two key frameworks operationalize these principles for contemporary practice. The OCAP principles---Ownership, Control, Access, and Possession---developed by the First Nations Information Governance Centre, establish that Indigenous communities must own, control, access, and possess data about their peoples~\cite{FNIGC-sp, Kukutai2016-nf}. These principles ensure information governance aligns with self-determination rather than external research agendas. The CARE Principles---Collective Benefit, Authority to Control, Responsibility, and Ethics---complement OCAP by emphasizing relational accountability and collective well-being over individual data rights~\cite{Carroll2020-mq}.

Scholars working at the intersection of Indigenous studies and disability studies have extended these sovereignty frameworks to critique mainstream accessibility discourse. Meissner and Reynolds argue that dominant conceptions of access in disability theory presuppose settler-colonial property relations, proposing ``deep access''---a framework rooted in kinship, reciprocal responsibility, and restoration of relationships severed by colonial violence~\cite{Meissner2024-tj}. Cowing introduces ``settler ableism'' to name how state assimilation measures target Indigenous bodyminds, arguing that ``occupied land is an access issue'' that feminist disability studies has largely failed to address~\cite{Cowing2020-on}. Empirical work reveals that many Indigenous peoples reject ``disability'' as an alien Western construct, instead celebrating uniqueness and community contribution~\cite{Rivas-Velarde2018-iu}; Moola et al. use disability justice to examine how colonialism produces and maintains Indigenous childhood disability through interlocking institutional failures~\cite{Moola2024-iw}. This scholarship establishes that relational, collective, and sovereignty-based framings of access reflect longstanding Indigenous epistemologies---not novel theoretical interventions.

% We believe these Indigenous studies offer HCI valuable models for thinking beyond individual autonomy toward collective governance, reciprocal relationships, and community-defined standards. For disability communities similarly navigating power asymmetries in technology design, Indigenous sovereignty frameworks provide concrete vocabularies for asserting definitional authority over technological futures.

We believe these frameworks offer HCI valuable models for thinking beyond individual autonomy toward collective governance, reciprocal relationships, and community-defined standards. For disability communities similarly navigating power asymmetries in technology design, Indigenous sovereignty frameworks---and the Indigenous disability studies scholarship that has already begun this translation---provide concrete vocabularies for asserting definitional authority over technological futures. We draw on these insights while recognizing, following Tuck and Yang~\cite{Tuck2012-bp}, that conceptual borrowing cannot substitute for the political project of decolonization itself.

\section*{Positionality and Reflexivity}
This work is conducted by multiply disabled, neurodivergent, and allied researchers. We acknowledge that our constructivist process was shaped by our lived experiences. Our interpretations center a Disability Justice perspective, prioritizing identity-first language~\cite{Andrews2022-mj} and interpretations that align with principles of collective access~\cite{Invalid2017-zi}. Our analysis is one situated interpretation, not a ``view from nowhere''~\cite{Haraway1988-qi}.

\section{Methodological Approach}
\label{sec:methods}
% To address the central question of how independence and interdependence are negotiated in social accessibility research, we employed a multi-stage analytical approach combining systematic corpus construction with theoretical triangulation. Our methodology sought to 
% explore unresolved tensions in how the field negotiates relational modes, without presupposing what form resolution might take.

Our search began with the goal of understanding how social accessibility research has evolved and what tensions characterize the field. For this, we employed a multi-stage analytical approach (shown in Figure~\ref{fig:method-phrases-diagram}). As our initial intention was exploratory, we began without predetermined frameworks, allowing patterns to emerge through systematic corpus construction and iterative analysis. Only after identifying persistent tensions from our constructivist coding~(§~\ref{subsec:theme-development-coding}) around independence and interdependence did we employ theoretical triangulation to interpret them~(§~\ref{sec:productive-tension}).

 % Our research trajectory evolved thro: First, we sought to map the landscape of social accessibility research through collecting a systematically constructed and theoretically rich corpus. Second, by analyzing this corpus, we identified recurring tensions that cut across multiple domains. Finally, we recognized that these tensions revealed a deeper issue about definitional authority that we came to call sovereignty. This evolution was not planned but emerged through iterative engagement with the data.

\begin{figure*}[tb]
    \centering
    \includegraphics[width=1.0\textwidth]{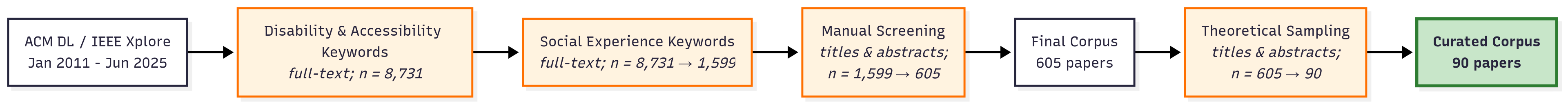}
    \caption{Illustration of process to obtain curated corpus for analysis.}
    \label{fig:method-stages}
    \Description{Flowchart of the study pipeline. From left to right: (1) Systematic dataset construction (n=605); (2) corpus curation (n=90); (3) inductive framework generation via constructivist grounded theory analysis; leading to (4) the resulting Three Praxis Framework. Arrows indicate a sequential progression from data collection to the final framework.}
\end{figure*}
% To understand how independence and interdependence were being considered in social accessibility research, we conducted a constructive grounded analysis~\cite{Charmaz2006-ty, Charmaz2017-av} of 90 theoretically rich papers on social accessibility. In order to (ensure we covered our bases) we  began by (1) constructed a comprehensive dataset of relevant literature with keyword searches, (2) used theoretical sampling to curate a theoretically rich corpus from the dataset. (TODO: add about additional thematic analysis process + meta-synthesis with lenses)

\subsection{Phase 1: Systematic Dataset Construction (n=605)}
\label{subsec:corpus-construction}

Our corpus construction (shown in Figure~\ref{fig:method-stages}) began with establishing temporal and topical boundaries. We selected January 2011 through June 2025 as our review period, anchored by Shinohara and Wobbrock's foundational articulation of social accessibility as a distinct research concern~\cite{Shinohara2011-bf}. This timeframe captures the field's evolution from primarily functional orientations toward more nuanced engagement with social and relational dimensions of assistive technology.

\subsubsection{Database Selection and Search Strategy}
\begin{figure}[ht]
    \centering
    \includegraphics[width=\columnwidth]{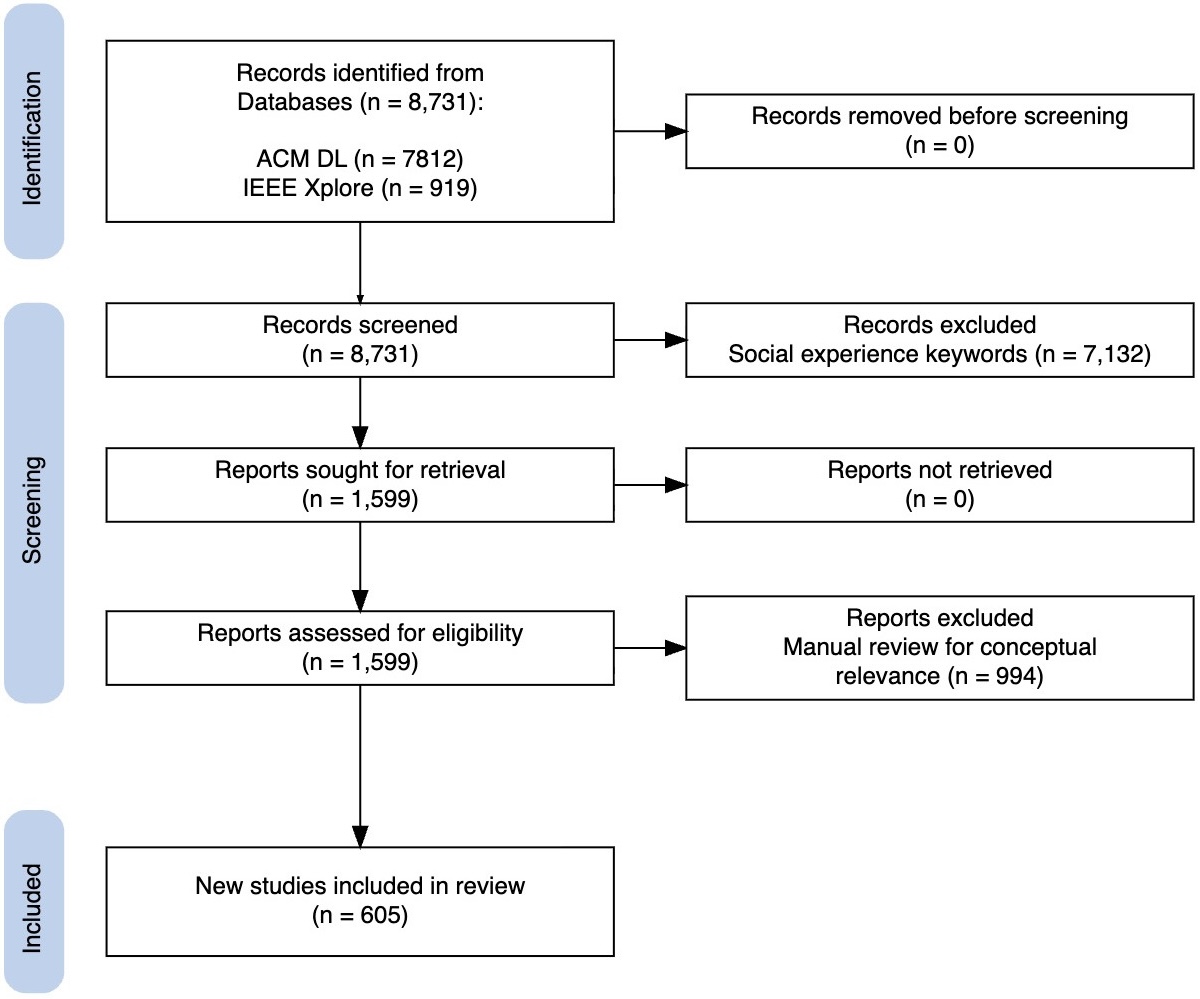}
    \caption{Steps of Systematic Dataset Construction}
    \label{fig:prisma}
    \Description{Horizontal flowchart of literature selection. Starting from ACM Digital Library and IEEE Xplore (Jan 2011–Jun 2025), searches with disability and accessibility keywords return n=8,731 full-text hits and 7812 and 919 for ACM DL and IEEE Xplore respectively. Applying social experience keywords narrows results to n=1,599. Manual screening of titles and abstracts reduces this to n=605, yielding a Final Corpus of 605 papers.}
\end{figure}

We targeted key HCI venues (ACM: ASSETS, CHI, CSCW, DIS, TACCESS, TOCHI, UIST; IEEE: HRI, ISMAR, VR, VIS), conducting full-text searches ACM and IEEE databases using the search terms listed in Figure~\ref{tab:keywords}. This two-stage protocol yielded 8,731 initial results, refined to 1,599 through social keyword filtering, and ultimately 605 papers through manual screening of titles and abstracts by the first author for conceptual relevance (see Figure~\ref{fig:prisma} for a PRISMA-style diagram).
% \vspace{1em}
\enlargethispage{\baselineskip}

% Search terms table
%\begin{table}[!h]
\noindent
\aptLtoX{\begin{table}[hbp!]
\renewcommand{\arraystretch}{1.1}
    \centering
    \small 
    \begin{tabular}{@{} l p{7.4cm} @{}}
    \toprule
    \multicolumn{2}{l}{\hspace*{-\tabcolsep}
    \cellcolor{gray!12}\textbf{\textcolor{black}{ACM Digital Library}}} \\
    \midrule
    \addlinespace[2pt]
    \textbf{Step 1} & \texttt{"disability" OR "accessibility" OR "assistive technology" OR "people with disabilities" OR "impairment"} \\
    \addlinespace[4pt]
    \textbf{Step 2} & \texttt{AND ("social acceptability" OR "stigma" OR "social perception" OR "communication" OR "collaboration" OR "identity" OR "belonging" OR "social norms" OR "masking" OR "expression")} \\
    \multicolumn{2}{l}{\hspace*{-\tabcolsep}\cellcolor{gray!12}\textbf{\textcolor{black}{IEEE Xplore}}} \\
    \addlinespace[2pt]
    \textbf{Step 1} & \texttt{("disability" OR "accessibility" OR "assistive technology" OR "people with disabilities" OR "impairment")} \\
    \addlinespace[4pt]
    \textbf{Step 2} & \texttt{AND ("social acceptability" OR "stigma" OR "social perception" OR "communication" OR "collaboration" OR "identity" OR "belonging" OR "social norms" OR "masking" OR "expression")} \\
    \bottomrule
    \end{tabular}
    \caption{Keyword search terms used for filtering steps}
    \label{tab:keywords}
\end{table}}{\begin{table}[hbp!]
\renewcommand{\arraystretch}{1.1}
    \centering
    \small 
    \begin{tabular}{@{} l p{7.4cm} @{}}
    \toprule
    % \textbf{Stage} & \textbf{Search Terms} \\
    % \midrule
    
    % --- Database 1 Header ---
    \rowcolor{gray!12}
    \multicolumn{2}{l}{\hspace*{-\tabcolsep}\textbf{ACM Digital Library}} \\
    \addlinespace[2pt]
    \textbf{Step 1} & \texttt{"disability" OR "accessibility" OR "assistive technology" OR "people with disabilities" OR "impairment"} \\
    \addlinespace[4pt]
    \textbf{Step 2} & \texttt{AND ("social acceptability" OR "stigma" OR "social perception" OR "communication" OR "collaboration" OR "identity" OR "belonging" OR "social norms" OR "masking" OR "expression")} \\
    
    \midrule
    
    % --- Database 2 Header ---
    \rowcolor{gray!12}
    \multicolumn{2}{l}{\hspace*{-\tabcolsep}\textbf{IEEE Xplore}} \\
    \addlinespace[2pt]
    \textbf{Step 1} & \texttt{("disability" OR "accessibility" OR "assistive technology" OR "people with disabilities" OR "impairment")} \\
    \addlinespace[4pt]
    \textbf{Step 2} & \texttt{AND ("social acceptability" OR "stigma" OR "social perception" OR "communication" OR "collaboration" OR "identity" OR "belonging" OR "social norms" OR "masking" OR "expression")} \\
    
    \bottomrule
    \end{tabular}
    \caption{Keyword search terms used for filtering steps}
    \label{tab:keywords}
\end{table}}
%\caption{Search Terms Used for Each Filtering Stage}
%\label{tab:keywords}
%\end{table}

\subsection{Phase 2: Theoretical Sampling and Corpus Curation (n=90)}
\label{subsec:curating-corpus}
From the comprehensive dataset, the first author employed theoretical sampling~\cite{Charmaz2017-av} to assemble a focused corpus by assessing each works' title and abstract (also shown in Figure~\ref{fig:method-stages}). (For a comprehensive bibliometric analysis of this corpus, including venue distributions and thematic territories, see \textit{The Three Praxes: A Framework and Map of Social Accessibility Research}~\cite{Jang2026ThreePraxes}.) This purposive selection prioritized papers offering rich conceptual insights into the independence--interdependence dynamic.

%\medskip
\vspace{1em}
\noindent \textbf{Inclusion criteria:}
\begin{itemize}[topsep=0em]
    \item[(a)] Explicit engagement with social or experiential dimensions of disability and technology (e.g., identity negotiation, stigma management, relational dynamics)
    \item[(b)] Employment of interpretive or critical methodologies providing thick description (e.g., ethnography, participatory design, phenomenological inquiry)
    \item[(c)] Direct theoretical engagement with relevant frameworks (e.g., disability studies, STS, critical theory)
\end{itemize}

\vspace{1em}
\noindent\textbf{Exclusion criteria:} 

\noindent Papers focusing exclusively on technical optimization or functional metrics without substantive engagement with social dimensions were excluded, as these would not contribute to theorizing the relational dynamics central to our inquiry.

\subsection{Phase 3: Three-Stage Thematic Analysis with Theoretical Triangulation}
\label{subsec:analysis}

Our analytical approach combined inductive pattern identification with deductive theoretical interpretation, allowing empirical insights and theoretical frameworks to inform each other dialectically.

\subsubsection{Stage 1: Constructivist Coding and Theme Development}
\label{subsec:theme-development-coding}
\begin{figure*}[t]
    \centering
    \includegraphics[width=1\textwidth]{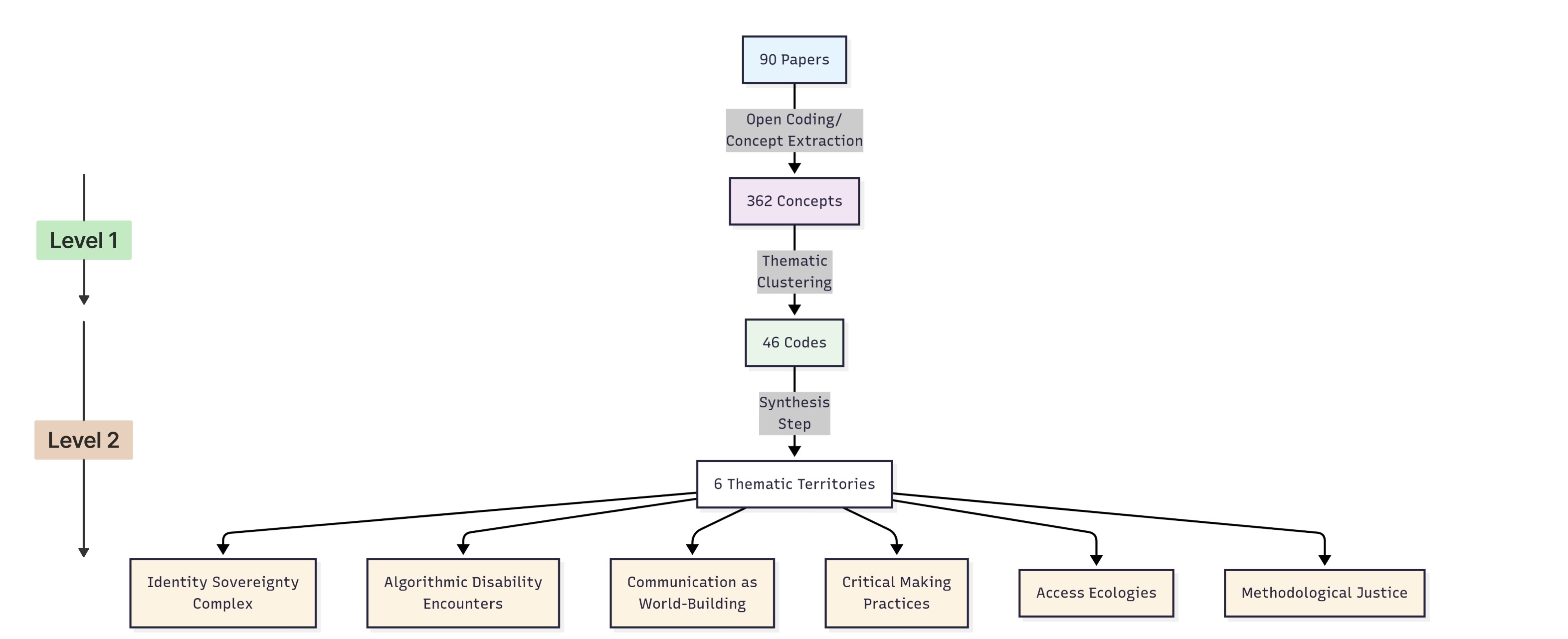}
    \caption{Illustration of the constructivist grounded theory analysis process (Stage 1): we move from conceptual extraction to thematic synthesis.}
    \label{fig:method-papers-to-themes}
    \Description{This diagram illustrates the multi-stage analytical process employed in this study, progressing from raw data to theoretical frameworks. Beginning with 90 papers, open coding and concept extraction yielded 362 distinct concepts. Through thematic clustering, these concepts were consolidated into 46 codes, which were further synthesized into six thematic territories: Identity Sovereignty Complex, Algorithmic Disability Encounters, Communication as World-Building, Critical Making Practices, Access Ecologies, and Methodological Justice.}
\end{figure*}

We followed constructivist grounded theory principles~\cite{Charmaz2006-ty} and situational analysis~\cite{Clarke2005-wk} treating papers, concepts, and infrastructures as elements in an analytic situation rather than neutral carriers of data~\cite{Coffey1996-mw, Prior2008-nz}. Beginning with an open coding of the 90 papers, we focused on how independence and interdependence were conceptualized, experienced, and negotiated. Using digital affinity diagramming, the first author extracted 362 discrete concepts from the full-text works, preserving where possible the original language to maintain semantic fidelity (e.g., ``algorithmic disclosure mediation,'' ``crip aesthetics of augmentation,'' ``platformed access labor'').

% Through constant comparative analysis, we developed:
% \begin{itemize}
%     \item \textbf{46 focused codes} through clustering semantically related concepts (e.g., ``strategic visibility management,'' ``augmentative identity work'')
%     \item \textbf{6 thematic categories} through axial coding examining relationships between codes:
%     \begin{itemize}
%         \item Identity Sovereignty Complex
%         \item Communication as World-Building
%         \item Access Ecologies
%         \item Critical Making Practices
%         \item Algorithmic Disability Encounters
%         \item Methodological Justice
%     \end{itemize}
% \end{itemize}

% \subsection{A Two-Stage Thematic Analysis}

% To move beyond simply documenting the independence-interdependence paradox toward theorizing its resolution, we employed a two-stage thematic analysis that combined inductive synthesis with theoretical interpretation. This approach allowed us to first surface empirical patterns from the data, then interpret these patterns through productive theoretical tensions.

% \paragraph{Stage 1: Inductive Thematic Synthesis}

% Our initial analysis employed constructivist grounded theory principles \cite{Charmaz2006-ty, Charmaz2017-av, Charmaz2017-cs}, acknowledging that theoretical insights emerge from the interaction between researchers' perspectives and empirical data. 
Each concept was captured as a discrete node with its source citation (e.g., ``dynamic disclosure toggles,'' ``celebratory aesthetics,'' ``access intimacy,'' ``platformed audism''). Through iterative axial coding, we analyzed relationships between these conceptual nodes, moving from a flat collection to a structured intellectual landscape.  This process is shown in Figure~\ref{fig:method-papers-to-themes} and involved two levels of synthesis:

\noindent \paragraph{\textbf{Level 1} -- Clustering:} We grouped semantically related concepts into 46 granular codes. For instance, concepts like ``masking as performance and suppression,'' ``strategic muteness as social filter,'' and ``double-edged digital masking'' were clustered under \textit{Masking, Passing, Accommodation}. Similarly, ``DIY-AT as systemic critique,'' ``replicative DIY-AT,'' and ``social justice-aligned makerspaces'' formed the code \textit{DIY-AT \& Making}.

\noindent\paragraph{\textbf{Level 2} -- Theme Synthesis:} These 46 codes were then synthesized into six overarching themes through analysis of their interrelationships:
\noindent \begin{itemize}
    \item \textbf{Identity Sovereignty Complex}: How disabled people negotiate disclosure, masking, and visibility in relation to identity construction
    \item \textbf{Communication as World-Building}: The role of social acts creating communicative worlds
    \item \textbf{Access Ecologies}: Networks of interdependence, care, and the distribution of access labor
    \item \textbf{Critical Making Practices}: DIY-AT, celebratory technologies, and resistance through material practice
    \item \textbf{Algorithmic Disability Encounters}: AI mediation, data representation, and algorithmic/digital identity construction
    \item \textbf{Methodological Justice}: Research ethics, co-design approaches, and epistemic considerations
\end{itemize}

\noindent\textbf{Identifying the Central Tension:} As we analyzed across these six themes, a recurring pattern was surfaced that cut through all of them. In the \textit{Identity Sovereignty Complex}, users wanted both privacy protection (independence) and community validation (interdependence). Within \textit{Access Ecologies}, the same support networks that provided essential care also created vulnerability to others' schedules and continued presence. The \textit{Critical Making Practices} territory revealed DIY-AT as simultaneously individual acts of self-determination and deeply reliant on collective knowledge-sharing infrastructures. Even in \textit{Communication as World-Building}, AAC users navigated between autonomous self-expression and dependence on communication partners' interpretive labor.

This cross-cutting pattern revealed what we term the \textit{Independence--Interdependence Complex}: a persistent, unresolved tension where disabled people simultaneously demanded greater individual control \textbf{\textit{and}} more robust collective support. Crucially, this was not a fixed binary choice---the same users, in the same contexts, articulated needs for both. The complexity lay in how these seemingly opposing orientations were actually co-constitutive: independence often required interdependent infrastructures to achieve, while meaningful interdependence demanded individual agency to direct.

\subsubsection{Stage 2: Theoretical Triangulation}
\label{sec:theoretical-triangulation}
To interpret the \textit{Independence–Interdependence Complex}, we employed a theoretical triangulation approach~\cite{Carter2014-yq,Denzin2015-ep}, involving iterative cycles of mapping, comparison, and reinterpretation.

\paragraph{Analytical Process:} The first author re-analyzed our 362 concepts with our six themes and the three lenses (The Self, The Interaction, The Body). For each intersection, we asked: How would this lens interpret these findings? What does it illuminate? What does it obscure? 

For the first lens (The Self), we examined how accounts spoke to needs for autonomy-as-volition, competence, and relatedness, and whether these appeared supported or thwarted. For The Interaction, our analysis followed the Chicago lineage~\cite{Mead1967-vq, Blumer1992-oz} and negotiated-order perspectives~\cite{Strauss1978-dq}, attending to how papers \textit{defined the situation} (problems, roles, audiences), mobilized stigma symbols and impression management~\cite{Goffman1959-uf}, and stabilized expectations through role ascriptions. Treating papers as social actors rather than neutral reports, we looked for symbolic objects (e.g., ``autonomy,'' ``burden''), audience design (clinician, caregiver, platform), and tactics such as covering, passing, and boundary work. This SI-inflected stance is consistent with our constructivist grounded theory methodology~\cite{Charmaz2006-ty, Charmaz2017-av, Charmaz2017-cs}, which likewise assumes that accounts are situated, interactional productions rather than transparent windows onto reality. For The Body, we looked at how technologies, infrastructures, and institutions configured human–nonhuman assemblages and whose norms and priorities those assemblages served.

\paragraph{Illustrative Example:} When we applied these lenses to the ``forced isolation'' experiences in our data, each illuminated different dimensions. SDT revealed these as violations of the need for relatedness. Crip technoscience reframed the same experiences as products of systemic biases that make independence compulsory. Symbolic interactionism showed how ``isolation'' itself was a negotiated meaning---what one person experienced as abandonment, another framed as finally being left alone.

\subsubsection{Stage 3: Synthesis through Productive Tension}
\label{sec:productive-tension}
The most generative insights emerged not from any single lens but from the incompatibilities between them.

\paragraph{Theoretical Conflicts:} SDT's universalist claims about human needs directly conflicted with crip technoscience's insistence that ``needs'' themselves are socially constructed and politically contested. When our corpus equated independence with autonomy, SDT (properly applied) revealed this as a conflation---volition could be satisfied through chosen interdependence. Crip technoscience went further, questioning whether ``autonomy'' itself encoded ableist norms about self-sufficiency. Symbolic interactionism complicated both by showing how ``autonomy'' meant different things in different interactional contexts---sometimes independence, sometimes the choice of interdependence, sometimes the power to define the interaction itself.

\paragraph{Adjudication Principles:} When lenses conflicted, we adopted three principles. First, we privileged situated accounts---participant aims and experiences---over theoretical priors. Second, we treated ``autonomy'' strictly as volition (following SDT's technical definition) and separately inspected whether that volition was recognized or constrained by others and infrastructures. Third, we used symbolic interactionism to disambiguate what ``autonomy'' or ``support'' meant \textit{in situ}, recognizing that identical terms carried different meanings across contexts.

\paragraph{Reading The Body Diffractively:} Within our third lens, we read posthumanism and crip technoscience diffractively rather than as a synthesis~\cite{Barad2009-qs, Geerts2016-wu}. Posthumanist perspectives highlighted ontological entanglement---the impossibility of separating human from nonhuman actors. Crip technoscience foregrounded governance and technoableism---asking whose bodies are centered and who controls the terms of technological mediation. Where entanglement and volition collided, we asked: who gets to set terms, revise goals, and seek recourse? 

\paragraph{Emergence of Sovereignty:} These adjudication moves yielded sovereignty as the through-line. Users were not simply seeking volition (SDT's autonomy) or challenging norms (crip technoscience's resistance) or negotiating meanings (symbolic interactionism). They were seeking something more fundamental: the \textit{recognized power to set the terms} of their relationships with both technology and people. This theoretical synthesis produced our core distinction between \textit{autonomy} and \textit{sovereignty}---the recognized power to choose one's relational mode and define the goals of technology use. This shift reframes the central question from ``Can they do it?'' to ``Do they get to decide?''

\subsection{Transparency and Reproducibility}
To support research transparency and enable future theoretical development, we make available our complete analytical materials: the full 605-paper corpus from our systematic search, the curated 90-paper theoretical sample, and the complete coding structure generated during analysis (362 conceptual nodes, 46 focused codes, and 6 theoretical categories). These materials, along with our coding memos and theoretical mapping matrices, are provided in the supplementary materials to facilitate both verification of our analysis and extension of the sovereignty framework to other contexts.

% \section{Findings: From Autonomy to Sovereignty}
% \label{sec:findings}
% We entered this research seeking to understand how the field conceptualizes social accessibility. Through our inductive synthesis, we discovered recurrent tensions that the independence-interdependence spectrum alone could not resolve. Our subsequent meta-synthesis with three theoretical lenses revealed that these tensions consistently turned on \textit{who sets the terms of relation and success}. This pattern led us to develop the concept of relational sovereignty: the recognized power to determine aims, boundaries, and dependencies, and to have those decisions upheld by others and surrounding infrastructures.

% This section presents our central theoretical contribution: \textit{relational sovereignty}. Drawing on the two-stage analysis, our inductive synthesis surfaced recurrent tensions that the independence–interdependence spectrum alone could not resolve, while our meta-synthesis showed that these tensions consistently turned on \textit{who sets the terms of relation and success}. We use sovereignty to name this definitional authority: the recognized power to determine aims, boundaries, and dependencies, and to have those decisions upheld by others and surrounding infrastructures.

\section{Findings: From Autonomy to Sovereignty}
\label{sec:findings}
We began this research seeking to understand how the field conceptualizes social accessibility. Through our inductive synthesis, we discovered recurrent tensions that the independence-interdependence spectrum alone could not resolve. Our subsequent meta-synthesis with three theoretical lenses revealed that these tensions consistently turned on \textit{who sets the terms of relation and success.} This pattern led us to develop the concept of relational sovereignty: the recognized power to determine aims, boundaries, and dependencies, and to have those decisions upheld by others and surrounding infrastructures.

\subsection{Autonomy in Social Accessibility}
% The persistent tensions in our corpus revealed fundamental limitations in how the field conceptualizes autonomy in assistive technology. The same phenomenon often appeared simultaneously as both empowering and constraining---a paradox that positioning along an independence-interdependence spectrum could not explain. The examples that follow are illustrative rather than exhaustive, highlighting particular works that offer detailed accounts of patterns observed throughout the corpus.
The persistent tensions in our corpus revealed fundamental limitations in how the field conceptualizes autonomy in assistive technology. Despite critical disability scholarship arguing that independence is ontologically a myth—a form of interdependence with infrastructure rather than people~\cite{Mingus2017-uu, Hamraie2017-cq}—HCI and AT research largely continues to treat independence as a truly distinct, achievable, and superior goal~\cite{Shinohara2012-gd, Shinohara2018-eo}. Papers routinely framed independence as the endpoint of successful AT intervention, obscuring reliance on non-human elements while foregrounding freedom from human assistance. This ideological commitment to functional independence sits in tension with the material reality of disabled people's lives, which are inherently interdependent. The same phenomenon often appeared simultaneously as both empowering and constraining---a paradox that positioning along an independence-interdependence spectrum could not fully explain. The examples that follow are illustrative rather than exhaustive, highlighting particular works that offer detailed accounts of patterns observed throughout the corpus.

\subsubsection{The Instability of Autonomy}
We found this tension manifest most clearly in how the field handles the concept of autonomy. While self-determination theory defines autonomy as volition-the ability to act according to one's own values and interests~\cite{Deci1985-so, Ryan2000-de}-our corpus revealed that autonomy was frequently equated with independence. Papers describing autonomous technology use often meant users operating devices without assistance, rather than users exercising choice over their relational mode.

For instance, Silva et al.~\cite{Silva2023-zi} documented how smartwatch systems designed to promote children's autonomy actually enforced independence by reducing parental involvement, even when children preferred collaborative support. Their deployment study revealed tensions where children appreciated how the smartwatch served as a persistent reminder and enabled them to track daily goals independently, yet the multi-device nature of the system (parent phone/child watch) created rigid divisions that limited flexibility in co-regulation approaches. The transition from co-regulation to self-regulation was assumed to be linear and unidirectional, ignoring users who chose sustained interdependence~\cite{Silva2023-zi}. 

Conversely, the same study showed how technology serving as a ``bad cop'' enabled a form of autonomy by allowing users to blame the device for unpopular decisions (e.g., screen time limits), preserving relationships while exercising choice---a form of autonomy achieved through strategic dependence on technology~\cite{Silva2023-zi}. Read through SDT, these cases reveal that ``independence'' framed as self‑regulation often masked \textit{needs frustration}. Children were steered away from co-regulation even when they valued it, undermining relatedness and, paradoxically, their sense of autonomy as volition. Conversely, arrangements in which users chose to rely on partners or devices-``autonomy-supportive interdependence''-satisfied both autonomy and relatedness.

\subsubsection{Interdependence as Both Freedom and Coercion}
Interdependence was also represented with ambivalence. On one hand, Bhattacharjee et al.~\cite{Bhattacharjee2019-sk} find that robot-assisted feeding systems can operate within rich ecosystems of care that provide relational support and enable community participation beyond simple functional assistance. Higgins et al.~\cite{Higgins2024-ly} model this interdependence as fundamental to the success of AT initiatives, demonstrating how government, university, and community stakeholders can collaborate to provide equitable access in a statewide AT program. On the other hand, reliance on these networks creates profound vulnerability. When formal support systems like insurance prove inflexible or inadequate, Higgins et al.~\cite{Higgins2024-ly} note that users are left without access to necessary devices. This vulnerability is acute during transitions of care, such as when a key individual champion for accessibility leaves an institution, risking the collapse of the support they single-handedly provided, as described by Coverdale et al.~\cite{Coverdale2024-gv}.

% The institutional dimension of this interdependence is particularly fraught. For instance, Higgins et al.~\cite{Higgins2023-jw, Higgins2025-rg} document how disabled students, finding university accommodation systems to be rigid and inadequate, may circumvent official channels to create their own networks of mutual aid, framing interdependence as a form of resistance. However, as Coverdale et al.~\cite{Coverdale2024-gv} reveal, these alternative networks are often precarious, depending on the tireless work of individual ‘heroes’ or ‘champions’ whose departure could cause the entire support system to collapse. Furthermore, the technologies themselves can introduce new dependencies and stakeholder conflicts. Bragg et al.~\cite{Bragg2021-sw} describe how ownership of community-generated data is not a simple concept but a complex bundle of legal, cultural, and monetary claims. This complexity, they argue, creates vulnerabilities to mission creep where data collected for one purpose is later exploited for another, revealing how dependencies can create conflicting interests among the technology users, creators, and funders~\cite{Bragg2021-sw}.

Platform-mediated interdependence is especially complex. Video conferencing and social VR platforms, for example, can enforce ableist communication norms. By privileging spoken interaction, their design can marginalize users who stutter, as found by Wu~\cite{Wu2023-su}, or users who do not use voice communication (``mutes''), as described by Chen et al.~\cite{Chen2025-xr}, creating structural barriers to participation. Simultaneously, these same platforms are celebrated for their ability to incubate new communities by overcoming geographical distance. For instance, Wu~\cite{Wu2023-su} highlights how videoconferencing has been vital for connecting the globally dispersed community of people who stutter, enabling support groups and fostering a sense of shared identity. This paradox highlights a central tension: the same technological interdependence that connects communities can also subject them to platform governance and design choices they do not control. As Chen et al.~\cite{Chen2025-xr} illustrate, this lack of sovereignty can feel like a form of digital colonization, where users must adapt to disabling environments rather than having environments that adapt to them.

\subsubsection{Power, or The Ability to Set the Terms}

Across cases, what mattered most was not whether people acted independently or interdependently, but the power or \textit{who set the terms} under which relationships, visibility, and support were possible. In remote meetings, for example, Li et al.~\cite{Li2024-zf} found that people who stutter (PWS) consistently framed disclosure as a matter of control. Participants proposed tools that let them decide if, when, and how to disclose—e.g., a self-disclosure badge with tailored educational messages for non‑stuttering listeners, personalized ``encouragement'' prompts, and controls over when such prompts appear. They also designed features to shape downstream consequences of disclosure (e.g., ``I’m not done'' indicators to hold the floor; queue indicators; cool‑off periods that limit rapid re‑entry by fluent speakers). These ideas explicitly seek to redistribute conversational authority and redefine whose speech sets the pace and norms of the meeting~\cite{Li2024-zf}. Read together, they advance a form of disclosure sovereignty: beyond privacy or transparency---situated control over the contexts, timing, and implications of being known.

Design was also used as an explicit lever to challenge and reallocate power. Li et al. also engaged PWS co‑designers who proposed features that slow fluent speakers, make stuttered speech visible and legitimate (e.g., literal transcription of disfluencies), and offer non‑verbal mechanisms to claim conversational turns—design moves that intentionally disrupt inherited norms of speed, ``efficiency,'' and fluency so that disabled speakers can set (and reset) the rules of the interaction~\cite{Li2024-zf}. 

Power also organized access to assistive technologies and accommodations. Higgins et al. found that in a university makerspace, students described how clinicians, documentation regimes, and institutional procedures functioned as gatekeepers. Participants emphasized that ``safe'' AT design was contingent on relationships with medical professionals; without these relationships some projects were deemed too ``risky,'' and students with fewer clinical ties were less likely to participate. Students additionally characterized the accommodations system as privileged and opaque, often requiring extensive self‑advocacy and producing avoidance or disengagement; they turned to the makerspace to subvert inflexible policies and prototype alternatives that met their needs~\cite{Higgins2023-jw}.

In industry-facing ecosystems for children’s AT, Smolansky et al. found that ``efficacy'' operated as access currency. Professionals described prolonged funding processes (e.g., months‑long insurance approvals), trial requirements, and denials when approaches were not judged ``evidence‑based.'' Companies responded by developing templates and training to help caregivers and clinicians assemble acceptable rationales, and they called for better data pipelines to document impact—less to learn from children’s trajectories than to satisfy external gatekeepers~\cite{Smolansky2024-vd}.

Algorithmic infrastructures introduced further asymmetries. Kaur et al., in documenting  challenges that disability advocates in India faced, reported that platform ranking, moderation practices, and engagement‑driven cultures made disability content simultaneously hypervisible to abuse and invisible to support: patronizing comments and harassment were common; advocacy posts struggled to gain traction; and creators self‑censored or fragmented identities to mitigate harm. Participants characterized these dynamics as technoableist, arguing that platform architectures and policies actively invisibilize disabled voices and reassign the costs of safety to those least resourced to bear them~\cite{Kaur2024-lu}.

Finally, emerging emotion‑AI systems concentrated discretion to deployers while distributing risk unevenly. In a U.S.‑representative survey, Andalibi et al. found that comfort with emotion‑AI was low across all domains and especially low for workplace, social media, and hiring scenarios. Disabled people and gender minorities were significantly less comfortable than others, underscoring differential vulnerability; perceived accuracy predicted comfort but did not overcome baseline concerns, pointing to the need for regulation that centers identity‑based risk rather than technical fixes alone~\cite{Andalibi2025-xg}.

These findings show that assistive and sociotechnical systems are governed by power over disclosure, validation, and distribution of risk. People sought sovereignty over when and how to be seen; clinicians, insurers, and institutions defined what ``counts'' as legitimate need; and platforms and AI deployers determined whose experiences circulate and whose are sidelined. Attending to power---or who sets the terms of relation---explains patterns the independence–interdependence frame cannot and points toward design and policy interventions that redistrubte control to those most affected~\cite{Higgins2023-jw,Li2024-zf,Smolansky2024-vd,Kaur2024-lu,Andalibi2025-xg}.

\subsection{What Each Lens Revealed}
Our meta-synthesis clarified why the independence–interdependence spectrum alone could not resolve the tensions in our corpus. Each lens contributed a different, necessary piece and, read together, pointed to sovereignty as definitional authority over aims, boundaries, and dependencies.

\subsubsection{The Self: Separating Volition from Independence} SDT helped us differentiate autonomy (volition) against the our corpus' use of independence. Across AAC studies, users exercised volition by flexibly choosing communication modes and configuring support, even when that meant relying on others or tools. ``Adaptation'' and ``Fulfillment'' emerged as primary value themes in Zolyomi et al.'s work---users switch devices and modalities by context (waterproof, wearable, desktop), prize flexible turn-taking across in-person/Zoom/text, and foreground expressive goals (humor, sarcasm, teasing) and identity over raw throughput~\cite{Zolyomi2025-uh}. The need for ``augmenting, not replacing, competencies'' alongside a ``regulatable safety net'' is emphasized by Curtis et al., who prioritize user-defined competence and safeguards rather than pushing toward normative fluency benchmarks~\cite{Curtis2024-tk}. AAC functions as a relational mediator with shared workload and topic‑contingent effort, as Dai et al. demonstrate; effectiveness is distributed across partners and situations, clarifying that chosen interdependence can preserve autonomy~\cite{Dai2022-zt}. Together, these findings anchor our use of SDT's autonomy (volition) distinct from independence, and motivate attention to competence by one's own standards.

\subsubsection{The Interaction: Meaning as Negotiated} SI explained how meanings of autonomy, support, and success are interactionally produced. ``Collective Communication Access'' integrates the transactional model of communication to treat access as co-created by all participants, as McDonnell \& Findlater articulate; their framework derives practical implications (e.g., engaging all interlocutors, attending to local norms) that exceed sender–receiver framings~\cite{McDonnell2024-nv}. Survivorship involves navigating ``societal, relational, and personal boundaries,'' with Ankrah et al. revealing selective disclosure to manage others' emotions, negotiated timing/audience of identity revelations, and the emotional labor of setting relational terms—all demonstrating that disclosure is a negotiated social act, not a binary state~\cite{A-Ankrah2022-yt}. Through their analysis of ``voice–avatar dissonance,'' Chen et al. expose how platform defaults (e.g., audio/video prioritization, representational mismatch) can impose identities and interactional expectations that users must continually renegotiate \cite{Chen2025-xr}. The ``conduit'' metaphor receives critique from Zieli\'{n}ski et al., who develop a coordination metaphor for communication, making explicit that meaning is co-constructed and that interactional control is shared rather than transmitted, a core SI tenet \cite{Zielinski2022-xj}.

\subsubsection{The Body: Power and Governance} This lens highlighted power and norms, foregrounding questions of governance of human–machine systems. In our corpus analysis, ``posthuman care'' emerged as care distributed across humans, devices, and environments in Bogdanova et al.'s work; they interrogate algorithmic perception/digital gaze---how systems read and render bodies—implicating whose bodies and tempos are centered in sociotechnical practice~\cite{Bogdanova2024-bi}. The ``accuracy-comfort-risk'' paradox and differential vulnerability to algorithmic harm surface in Andalibi et al.'s analysis, demonstrating that technical gains can raise exposure risks or discomfort for disabled users if users lack authority over trade-offs~\cite{Andalibi2025-xg}. Inductively mapping ``axes of human‑like identity'' in HRI, Miranda et al. critique the pathologization of neurodivergence while noting glaring omissions (e.g., race), underscoring that identity ascription in sociotechnical systems is political and normatively loaded~\cite{Miranda2023-vl}. These contributions collectively justify shifting analysis from ``does it work?'' to ``\textit{who} sets the terms by which it works,\textit{ for whom}, and \textit{with what recourse?}''

Read across these lenses, we found a recurrent pattern: users sought not only volition (autonomy), but also desired authority to set the terms of relation and success. We use relational sovereignty to name that authority.

\enlargethispage{2\baselineskip}
\subsection{Parallels to Indigenous/Data Sovereignty Studies}
We adopt sovereignty mindful of its established meanings in Indigenous studies (collective self-determination against external governance) and data sovereignty (community governance over collection, interpretation, and circulation). Indigenous data governance frameworks like OCAP and CARE assert community authority over collection, interpretation, and circulation of information~\cite{Kukutai2016-nf, Carroll2020-mq}. Indigenous disability scholars have already argued that such relational, sovereignty-based framings better capture disabled experience than individualist access models~\cite{Meissner2024-tj, Cowing2020-on}. Our corpus analysis surfaces HCI work that moves toward similar commitments:

\textbf{Collective access as governance.} McDonnell \& Findlater synthesize disability justice with communication theory to argue that ``access is something that happens between people,'' recommending designs that ``engage all communication participants''~\cite{McDonnell2024-nv}. Similarly, Theil et al.'s deafblind participant explicitly ``imagined multiple devices to support social interactions for everybody,'' distributing inclusion work across all present~\cite{Theil2020-wj}. These enact governance as collective capacity—the group claims ownership and control over access norms (OCAP: Ownership/Control; CARE: Collective Benefit, Authority to Control).

\textbf{Redistributing interactional power.} Li et al. surface features that ``proactively disrupt existing communication flows to redistribute the power'' between speakers, including self-disclosure badges, floor-holding indicators, and enforced cool-off windows~\cite{Li2024-zf}. This resembles Indigenous jurisdiction—communities set and enforce protocols governing participation (CARE: Authority to Control; OCAP: Control/Access).

\textbf{Institutional legitimation of rest.} Janicki et al. reframe rest as access for chronic illness, using infrastructural speculations to surface where rest is legitimized or denied~\cite{Janicki2025-rt}. This mirrors sovereignty at institutional scales—establishing and governing supportive infrastructures (CARE: Collective Benefit, Responsibility; OCAP: Possession as custodial control).

\textbf{Disabled leadership in research.} Sum et al. position disability justice to center ``the needs and expertise of disabled people,'' calling for disabled people to ``lead the way''~\cite{Sum2022-lp}. Harrington et al. reinforce community-defined standards and accountability~\cite{Harrington2023-ct}—resonating with OCAP's Ownership/Control and CARE's Authority/Ethics.

These strands operationalize sovereignty as authority to define aims, boundaries, and dependencies and to have those definitions recognized and upheld by others and infrastructures.

\subsection{Autonomy versus Sovereignty}
The distinction between autonomy and sovereignty becomes clear when examining specific domains:

\textbf{Interactional terms.} Autonomy would mean ``I get a chance to speak.'' Sovereignty goes further: the group and platform co-enforce fair turn-taking, repair, and floor-holding so the speaker's terms actually stand. Zieli\'{n}ski et al.'s coordination metaphor shows how meaning is co-constructed~\cite{Zielinski2022-xj}, while Li et al.'s co-designed features explicitly redistribute conversational authority through ``I'm not done'' indicators and cool-off periods~\cite{Li2024-zf}.

\textbf{Disclosure and representation.} Autonomy would offer a toggle to disclose. Sovereignty is the ongoing authority to decide when, how, and to whom to disclose---and to have platforms align with those choices across contexts. Ankrah et al. reveal disclosure as layered boundary work~\cite{A-Ankrah2022-yt}, while Chen et al. show how platform defaults impose identities users must continually renegotiate~\cite{Chen2025-xr}.

\textbf{Pacing and competence.} Autonomy would say ``I can choose to pause.'' Sovereignty asserts a recognized right to slow down backed by spaces, policies, and tools that make pacing materially possible~\cite{Janicki2025-rt}. Similarly, sovereignty centers competence ``on my terms,'' with systems that augment strengths and value expressivity over normalization~\cite{Curtis2024-tk,Zolyomi2025-uh}.

\textbf{Institutional and algorithmic governance.} Autonomy would let users request services. Sovereignty demands reduced gatekeeping, reliable support across handoffs, and user control over algorithmic exposure. Smolansky et al. document ``efficacy as currency'' gatekeeping~\cite{Smolansky2024-vd}, while Andalibi et al. reveal how disabled users face differential vulnerability to algorithmic harm without authority over comfort-risk trade-offs~\cite{Andalibi2025-xg}.

\subsection{Why Sovereignty was Ultimately Necessary}
Across our three lenses, sovereignty emerged as the missing dimension. Self-determination theory clarifies volition~\cite{Deci1985-so,Ryan2000-de} but leaves power implicit when that volition meets entrenched norms. Symbolic interactionism reveals how norms are negotiated, yet remains agnostic about who should set them~\cite{Goffman1959-uf}. Crip perspectives press the governance question but need concrete vocabularies for collective authority~\cite{Hamraie2019-fx}. Indigenous and data-sovereignty frameworks supply these vocabularies through OCAP and CARE principles~\cite{Kukutai2016-nf,Carroll2020-mq}.

Our corpus demonstrates this need repeatedly: group-governed access reframes communication as collective capacity~\cite{McDonnell2024-nv}; co-designed features redistribute interactional power~\cite{Li2024-zf}; rest becomes institutionally legitimized~\cite{Janicki2025-rt}; and competence is user-defined rather than normalized~\cite{Curtis2024-tk}. The practical question shifts from ``Can they do it?'' to ``Do they get to decide---and will those decisions be recognized?''

\begin{figure*}[t]
    \centering
    \includegraphics[width=0.92\textwidth]{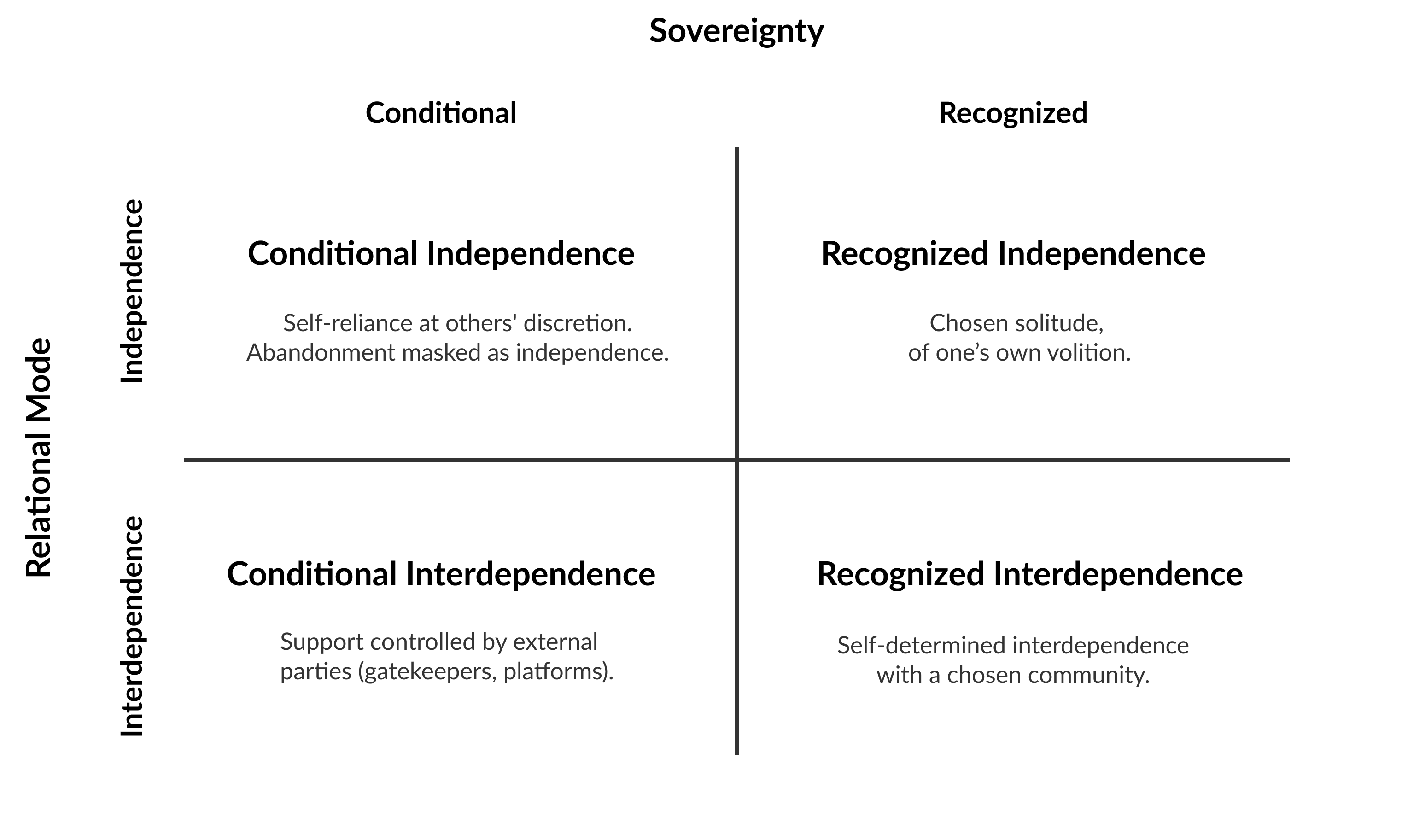}
    \caption{The Relational Sovereignty Matrix: The left column (Conditional) captures what is traditionally experienced as \textit{dependence}---a lack of power to set one's own terms. The goal of sovereignty-oriented design is movement rightward (toward Recognized Sovereignty).}
    \label{fig:relational-sov-matrix}
    \Description{A 2 by 2 matrix categorizing relational sovereignty based on two dimensions: Relational Mode (Independence vs Interdependence) and Sovereignty (Conditional vs Recognized). The four quadrants show: Conditional Isolation (conditional independence), Chosen Solitude (recognized independence), Conditional Reliance (conditional interdependence), and Chosen Community (recognized interdependence). Each cell includes brief explanatory text describing the nature of autonomy and support in that category.}
\end{figure*}
\enlargethispage{\baselineskip}
\section{The Theory of Relational Sovereignty}
\label{sec:theory-of-sovereignty}

We propose \textit{relational sovereignty} as a framework that resolves the independence--interdependence tension by introducing an orthogonal dimension of power (shown in Figure~\ref{fig:relational-sov-matrix}). Relational sovereignty refers to the recognized authority to choose one's relational mode---whether to act independently or interdependently---and to define the conditions under which those relationships operate. This shifts the central question from ``Can they do it alone?'' to ``Do they get to decide?''

This concept emerged from observing that similar relational arrangements could feel either empowering or oppressive depending on whether users had meaningful authority over the terms. The same support relationship can represent either chosen care or coerced dependence; the same platform can enable community or enforce isolation. These experiential differences cannot be explained by the relational mode alone---they require attention to who holds definitional power over aims, boundaries, and dependencies.

\subsection{The Relational Sovereignty Matrix}

To analyze agency beyond the independence--interdependence spectrum, we introduce a $2 \times 2$ matrix with orthogonal axes: (1) Relational Mode, from independence to interdependence; and (2) Sovereignty, from Conditional to Recognized. Crucially, we employ \textit{independence} here not as an ontological reality---acknowledging the critical theorist view that total self-reliance is a myth~\cite{Mingus2017-uu}---but as a descriptive category for the \textit{relational mode} of operating without direct human assistance. We present the Relational Sovereignty Matrix visually in Figure~\ref{fig:relational-sov-matrix}.

\subsubsection{Populating the Matrix: Four States of Being}

The intersection of the two axes generates four distinct experiential states. This structure shows that the ``what'' of the interaction (acting alone or with others) is less determinative of user experience than the ``how'' (under whose authority). By distinguishing these quadrants, we can see that true empowerment does not lie in moving from the bottom of the matrix to the top (from interdependence to independence), but in moving from the left to the right (from conditional to recognized sovereignty). Each quadrant is provided an example from our corpus in Table~\ref{tab:sovereignty-matrix-vignettes}.

\paragraph{Conditional Independence} 
Nominal independence becomes abandonment when supports lapse, platforms misalign, or institutions externalize responsibility: support discontinuities in assistive ecosystems place the burden back on the individual~\cite{Scougal2023-lr}; platform defaults such as voice–avatar misalignment and audist norms isolate users even when ``acting alone''~\cite{Chen2025-xr}; and gatekeeping or ``efficacy as currency'' privatizes access labor~\cite{Smolansky2024-vd}. Strategic invisibility, while protective, can further entrench isolation when settings do not back user choices~\cite{Sannon2023-np}. 

\paragraph{Conditional Interdependence} 
Interdependence becomes coercive when terms are set by others. Institutional and clinical gatekeeping mandates reliance under external standards~\cite{Smolansky2024-vd}; partner-skill dependencies and brittle handoffs tie success to others' availability~\cite{Scougal2023-lr}; and algorithmic logics produce accuracy–comfort–risk trade-offs that the user cannot refuse or reshape~\cite{Andalibi2025-xg}. Interactional norms that privilege speed and fluency likewise require people to adapt to others' pace absent group adjustments~\cite{Li2024-zf}.

\paragraph{Recognized Independence} 
Here, self-reliance is willed and reversible, backed by supports that keep doors open. Rest and pacing are recognized as access, not deficits, and are materially enabled by spaces, policies, and tools~\cite{Janicki2025-rt}; present-moment pacing is treated as a legitimate outcome rather than a failure to keep up~\cite{Thoolen2020-dw}; and ongoing boundary work and identity toggling give people authority over when and how to engage~\cite{A-Ankrah2022-yt,Wang2025-ql}.

\paragraph{Recognized Interdependence} 
Interdependence is sovereign when groups co-define and uphold accessible norms. Collective access reframes responsibility as shared and makes accessibility a group practice rather than an individual add-on~\cite{McDonnell2022-hr,McDonnell2024-nv}; co-designed controls (e.g., disclosure prompts, floor-holding signals, and pacing rules) redistribute interactional power so speakers' terms stand~\cite{Li2024-zf}; and disabled leadership orients goals and methods toward community-defined ends with accountable standards~\cite{Sum2022-lp,Harrington2023-ct}. In co-design with deafblind participants, sharing devices and obligations across all parties similarly relocates the work of inclusion onto the collective~\cite{Theil2020-wj}.

\begin{table*}[h]
\centering
\renewcommand{\arraystretch}{1.4}
\begin{tabular}{p{2.2cm} p{6.3cm} p{6.3cm}}
\toprule
& \textbf{Conditional} & \textbf{Recognized} \\
\midrule
\textbf{Independence} 
& \textbf{Conditional Independence} \newline
Chen et al.~\cite{Chen2025-xr} document how ``mutes'' in social VR face platform defaults that privilege spoken interaction. Though technically ``acting alone,'' these users experience isolation imposed by design choices they cannot negotiate, forcing adaptation to disabling environments rather than environments adapting to them.

& \textbf{Recognized Independence} \newline
Janicki et al.~\cite{Janicki2025-rt} reframe rest as access for people with chronic illness. When institutions provide quiet spaces, flexible pacing policies, and lending of restful artifacts, users can choose to step back without forfeiting future support. Solitude becomes willed and reversible as opposed to abandonment. \\

\midrule
\textbf{Interdependence} 
& \textbf{Conditional Interdependence} \newline
Smolansky et al.~\cite{Smolansky2024-vd} find that in children's AT ecosystems, ``efficacy'' operates as access currency. Caregivers and clinicians must navigate months-long insurance approvals, trial requirements, and denials when approaches are not judged ``evidence-based''---reliance on support whose terms are dictated by external gatekeepers.

& \textbf{Recognized Interdependence} \newline
Li et al.~\cite{Li2024-zf} co-designed videoconferencing features with people who stutter: self-disclosure badges paired with audience guidance, ``I'm not done'' indicators to hold the floor, and cool-off periods limiting rapid re-entry by fluent speakers. These redistribute conversational authority so that the group upholds the speaker's terms. \\

\bottomrule
\end{tabular}
\caption{Populating the Relational Sovereignty Matrix with vignettes from our corpus.}
\label{tab:sovereignty-matrix-vignettes}
\end{table*}

Taken together, the matrix keeps relational mode and power distinct and clarifies why similar surface arrangements (“getting help,” “doing it alone”) feel so different in practice. The decisive factor is governance: who sets the terms, under what standards, and with what recourse when terms are violated.

This framing also clarifies the status of \textit{dependence}. In prior literature, dependence is often treated as a third relational mode alongside independence and interdependence-the undesirable pole of a spectrum. Our analysis suggests otherwise: dependence is not a relational mode at all, but a sovereignty condition, or in other words, a condition of lacking power to actualize one's aims. It occurs when the terms of interaction---whether relying on a smart system, a human caregiver, or built infrastructure---are dictated by external forces without the user's ability to negotiate or refuse them. The entire left column of the matrix (Conditional Independence and Conditional Interdependence) captures what is traditionally experienced as dependence. Thus, the goal of sovereignty-oriented design is not to move users from dependence to independence, but from conditional to recognized sovereignty (rightward)---enabling chosen relational modes rather than prescribing which mode is correct.

\subsection{Telic Sovereignty: The Power to Decide the Terms}
\label{subsec:telic-sovereignty}
Beyond configuring who relates to whom and on what terms, sovereignty also concerns the ends of action. We call this \textit{telic sovereignty}: the recognized authority to define what counts as success in a given setting, to revise that telos (goals) across contexts and over time, and to have those definitions supported. 

Across the corpus, telic sovereignty appears as three consistent shifts. 
\begin{enumerate}
    \item From productivity to user-defined value: rest and pacing are treated as legitimate outcomes in their own right and are backed by spaces, policies, and tools that make slowing down possible~\cite{Janicki2025-rt,Thoolen2020-dw}. 
    \item From repair to augmentation and expression: competence is set “on my terms,” favoring augmentation over replacement and valuing expressive fulfillment alongside performance~\cite{Curtis2024-tk,Zolyomi2025-uh}.
    \item From individual metrics to collective obligations: success is relocated to whether groups enact and uphold accessible norms, not whether individuals cope—an orientation visible in collective-access programs and co-designed meeting features that redistribute interactional power~\cite{McDonnell2022-hr,McDonnell2024-nv,Li2024-zf,Theil2020-wj}.
\end{enumerate}

Telic sovereignty also reframes sociotechnical risk and identity; accuracy and visibility become conditional on user aims such as safety and comfort, with authority to govern exposure and audience rather than deferring to system defaults~\cite{Andalibi2025-xg,Sannon2023-np,Wang2025-ql}.

\section{Discussion: Implications and Proposed Interventions}
\label{sec:implcations}
The sovereignty framework was not our starting point but our destination---emerging from patterns we could not ignore in our systematic exploration of social accessibility research. It provides a framework for understanding the complex relationships between autonomy and interdependence in assistive technology. By examining who shapes the aims, boundaries, and dependencies of these systems—and how such decisions are supported—we can work toward more equitable designs. 

\paragraph{Situating Sovereignty in Crip and Critical Access Discourse}
We note that the concept of relational sovereignty does not come from a vacuum. The politics of interdependence articulated in Hamraie's \textit{Crip Technoscience Manifesto}~\cite{Hamraie2019-fx} already implies the need for disabled people to choose when, how, and on whom to rely. Mingus's concept of access intimacy names chosen, directed interdependence as a political and relational achievement~\cite{Mingus2017-uu}. Our contribution is not to invent these commitments but to crystallize and operationalize them for HCI practice.

Specifically, relational sovereignty offers three clarifications. First, it makes explicit that sovereignty is distinct from relational mode, allowing us to see that both independence and interdependence can be either chosen or coerced. Crip theory has long critiqued compulsory independence; our analysis and matrix extends this to show that interdependence, too, can be imposed rather than chosen (Conditional Reliance). Second, the framework gives designers and researchers a common language for connecting psychological needs (what does the user want?), social dynamics (how are meanings negotiated?), and political concerns (who holds power?). Third, the matrix offers a practical reframing: rather than asking whether a design supports independence or interdependence, we can ask whether it moves users toward having their choices recognized and upheld.

We also note parallels with critical access studies, which reframes access not as a technical problem to be solved but as an ongoing, political practice~\cite{Titchkosky2011-tg, Hamraie2017-cq}. Titchkosky's insight that access is ``something to think with'' rather than  satisfy aligns with our emphasis on who defines the terms of access. Hamraie's historical analysis of universal design reveals how access has always been shaped by power relations~\cite{Hamraie2017-cq}. Relational sovereignty extends this critical orientation into design and technical practice by making governance explicit: who sets terms, who can revise them, and what happens when terms are violated. This remainder of the section explores how sovereignty might inform social accessibility's theoretical models, design methods, and practical approaches.

\subsection{Reframing SDT Through Sovereignty}
Self-determination theory offers valuable insights into human needs, and when considered through the lens of sovereignty, its three components gain additional nuance that can guide AT design. Attending to relatedness complicates common HCI readings of SDT that equate autonomy with doing things alone. In our corpus, technologies or policies that pushed users toward self‑reliance (independence $\times$ conditional) regardless of their preferences often thwarted the need for relatedness, even when they appeared to support autonomy on the surface.

By contrast, many accounts of a chosen community (interdependence $\times$ recognized)---such as co-regulated AAC use or collaboratively negotiated access arrangements---simultaneously supported autonomy (volition over how to act) and relatedness (secure connection and belonging). This aligns with recent HCI syntheses of SDT that caution against isolating autonomy from the other needs or conflating it with independence~\cite{Tyack2024-gj}, and instead advocate a holistic interpretation in which autonomy-supportive interdependence can satisfy multiple needs at once.

\paragraph{Autonomy as Sovereignty} Moving beyond autonomy as individual execution, sovereignty emphasizes the importance of recognized authority over interactional terms. When videoconferencing platforms incorporate user-defined speaking protocols~\cite{Li2024-zf} or communities adopt collective access as shared practice~\cite{McDonnell2024-nv}, individual volition is upheld because others and surrounding infrastructures commit to enforcing the user’s terms. 

From an OIT perspective, many of the AT designed to provide autonomy in our corpus were in fact controlled: users acted to satisfy external demands (clinicians, insurers, institutional policies) rather than integrated, self-endorsed goals. Sovereignty makes this distinction visible by asking not only whether people follow a behavior, but whether they authored, can revise, and can refuse the terms under which that behavior counts as ``success.''

\paragraph{Competence on One's Own Terms} Rather than conforming to external benchmarks, sovereignty suggests that competence means achieving personally meaningful goals: whether expressive, social, or functional~\cite{Curtis2024-tk,Zolyomi2025-uh}. This perspective invites us to consider how systems might better support diverse forms of excellence, including forms that do not map neatly onto normative performance metrics.

\paragraph{Relatedness as Chosen Connection} Sovereignty recognizes that meaningful connection emerges when individuals can shape, limit, or refuse their engagement with others. This includes the capacity to negotiate, modify, or refuse relational obligations while maintaining respect for boundaries~\cite{A-Ankrah2022-yt,Chen2025-xr}. Our corpus makes clear that Conditional Independence often thwarts relatedness, leaving people isolated even the functional goals of self-reliance are met. By contrast, Recognized Interdependence can satisfy both autonomy (volition) and relatedness, aligning with SDT’s account of needs co-satisfaction.

Taken together, these reframings suggest new evaluation approaches: measuring successful renegotiations of relational terms, assessing how well groups and systems support user-defined norms, and examining alignment with individuals’ own goals rather than with imposed standards.

\subsection{Design Questions for Sovereignty-Oriented Practice}

We propose reflective questions that design teams might consider throughout their process, recognizing that context and communities will shape how these are addressed:

\begin{itemize}
    \item \textbf{Goals and Values}: How might systems honor users' diverse definitions of success---whether presence, expression, or rest---while remaining adaptable across contexts? (Consider~\cite{Janicki2025-rt,Thoolen2020-dw})
    
    \item \textbf{Interaction Design}: What opportunities exist for users to shape communication patterns, and how might communities or settings support these preferences? (Consider~\cite{Li2024-zf,McDonnell2024-nv})
    
    \item \textbf{Identity and Disclosure}: How can systems respect varied approaches to self-presentation and disclosure across different contexts? (Consider~\cite{A-Ankrah2022-yt,Wang2025-ql})
    
    \item \textbf{Distributed Support}: How might access labor become a shared opportunity for community building rather than an individual burden? (Consider~\cite{McDonnell2022-hr,Theil2020-wj})
    
    \item \textbf{System Governance}: Where might users benefit from greater control over algorithmic decisions, and how can we make these controls accessible? (Consider~\cite{Andalibi2025-xg,Sannon2023-np})
    
    \item \textbf{Equity and Recourse}: How can we ensure that diverse perspectives shape our designs and that meaningful recourse exists when issues arise? (Consider~\cite{Harrington2023-ct,Smolansky2024-vd})
\end{itemize}

\subsection{Building with Modular, Sovereign Components}
We advocate for assistive technologies composed of interoperable, user-controlled components that can be assembled, reconfigured, and integrated with existing digital infrastructures. Rather than black-box systems, this approach suggests collections of \textit{sovereign primitives}---modular capabilities that preserve ongoing choice and adaptability. This is in contrast to many traditional assistive technologies, which often provide ``on-rails'' experiences: with predetermined workflows with limited opportunity for adaptation or renegotiation. Once deployed, these systems may lock users into specific interaction patterns, making it difficult to adjust when needs change or contexts shift. This approach, while potentially ensuring consistency, can inadvertently strip users of the very agency that sovereignty seeks to preserve.

In contrast, consider how sovereignty might manifest through modular, reconfigurable components. The videoconferencing controls proposed by Li et al.~\cite{Li2024-zf} exemplify this approach: self-disclosure badges with customizable messages, ``I'm-not-done'' indicators to maintain speaking turns, and enforced cool-off periods to prevent interruption. Instead of being locked into a single platform's implementation, these could function as sovereign primitives that users deploy across different communication contexts---Zoom meetings, Discord channels, or workplace collaboration tools. A user might activate disclosure controls in professional settings while using different configurations for social interactions, adjusting parameters based on audience and context rather than being constrained to one-size-fits-all solutions.

Similarly, AAC users in Zolyomi et al.'s work~\cite{Zolyomi2025-uh} demonstrate sovereignty by fluidly switching between devices and modalities---waterproof options for swimming, wearable devices for mobility, desktop systems for extended composition. This multiplicity represents sovereignty in practice: not a single ``solution'' but an ecology of tools that users orchestrate according to their changing needs and contexts. By taking advantage of approaches like these, we enable disabled users to enhance the tools they already use rather than forcing migration to new platforms. This approach acknowledges that sovereignty includes the right to iterate---to experiment, adjust, and evolve one's technological supports without starting from scratch each time needs change.

\subsection{Considering Power in AT Design}
To attend to power in AT design, we must bridge the gap between capacity and governance. Schneider et al.~\cite{Schneider2018-us} characterize empowerment in HCI as split between 'power-to' (generative ability) and 'power-over' (relational authority). However, our analysis suggests that in the context of disability, these paths must converge: a user's 'power-to' is brittle without 'power-over' the terms of that action. Furthermore, sovereignty adds a new dimension to Schneider et al.'s psychological components of feeling, knowing, and doing: the meta-component of \textit{defining}. Through telic sovereignty~(§~\ref{subsec:telic-sovereignty}) we assert that true power encompasses not only the ability to complete a task, but also \textbf{\textit{the authority to define}} whether that task is worth doing at all.

Translating this authority into practice requires treating power distributions as explicit design materials. Rather than leaving governance opaque, designers must map decision points to reveal where users hold agency and where they face structural constraints. We believe this requires action beyond what many call system transparency; it involves surfacing hidden labor and establishing clear pathways for recourse when the system's definitions clash with the user's lived reality. By making these governance structures visible and manipulable, we transform the ``system'' from a black box into a negotiated space where expectations and responsibilities are mutually understood.

\subsubsection{For Researchers}
The sovereignty framework invites researchers to examine not only whether technologies function effectively, but also who defines effectiveness and whether those definitions can evolve with users' needs. This perspective encourages reflexivity about research methods themselves: whose questions guide our investigations, whose experiences constitute valid evidence, and how lived expertise is valued alongside other forms of knowledge~\cite{Ymous2020-gu}. Such reflexivity can strengthen research by broadening our understanding of what constitutes success and ensuring that diverse perspectives inform our work from conception through evaluation.

\subsubsection{For Disabled Users}
For disabled technology users, attention to sovereignty transforms assistive technology from accommodation within existing structures to tools that support self-determination. Rather than navigating systems that require constant negotiation for basic access, users gain recognized authority to shape their technological experiences. This shift makes access labor visible and potentially shareable across communities, reducing individual burden while building collective capacity. When sovereignty is embedded in design, it becomes possible to enforce one's preferences through technical and social mechanisms rather than repeated requests for accommodation. We know that each person's sovereignty will manifest differently---some preferring independence, others interdependence, many moving fluidly between modes as contexts change. Technology that respects this diversity supports not just task completion, but also dignity and authentic choice.

\section{Limitations}
\label{sec:limitations}

Our analysis represents one situated interpretation of social accessibility research, shaped by specific methodological choices. We examined English-language publications from premier HCI venues (ACM/IEEE) between 2011–2025, enabling systematic analysis within a coherent research community while necessarily excluding work from disability studies, STS, regional venues, and non-English outlets. This boundary provides a baseline understanding of how mainstream HCI has engaged with social accessibility, though the patterns we identify may be applicable beyond these venues. Our curation prioritized papers with explicit social or theoretical contributions in titles and abstracts. While this may underrepresent research embedding social considerations within technical contributions, it allowed focused analysis of how the field explicitly engages these concepts. 

Additionally, our meta-synthesis through our bodies of theories represents one analytical path among many---different theoretical lenses might surface complementary insights, which we view as promising for future work. The corpus reflects perspectives predominantly from Global North institutions, and sovereignty itself carries different meanings across cultural contexts. Our treatment of sovereignty would likely benefit from engagement with diverse cultural understandings of autonomy and authority, particularly from deeper Indigenous and Global South perspectives where sovereignty has distinct significance. These boundaries define our particular contribution: a systematic examination of how a specific research community has engaged with social accessibility over a formative period. We present the concept of relational sovereignty as a foundation for more generative investigations and critical engagement across the field.

% \section{Future Work}
% legal implications for sovereignty?

\section{Conclusion}
This paper introduces relational sovereignty as a framework for resolving the independence-interdependence tension in assistive technology. Through analysis of 90 social accessibility papers and theoretical synthesis, we identified that what matters is not whether someone acts alone or with support, but whether they have recognized authority to choose that state, define their goals and have their choices recognized by people, platforms, and institutions. We believe the shift from asking ``Can they do it?'' to ``Do they get to decide?'' can be illuminating for research in assistive technology design.

We introduce the Relational Sovereignty Matrix to operationalize sovereignty. It distinguishes between relational modes (independence/interdependence) and sovereignty (conditional/recognized), revealing four experiential quadrants. These distinctions explain why similar arrangements feel empowering in some contexts and oppressive in others. We observe through our analysis that the decisive factor often hinges on who sets the terms and whether recourse exists. We offer practical pathways through reframed motivational models, generative design questions, ideas for modular technical components, and a call for explicit attention to power in assistive technology design.

Relational sovereignty opens new research directions: how it operates at group rather than individual scales, how it translates across cultural contexts where autonomy and authority carry different meanings, and how evaluation methods might assess definitional power alongside functional outcomes. By making power and governance explicit design materials, we move toward systems where disabled people's aims are not merely accommodated but recognized and upheld. We hope this work provides a principled way to honor the diversity of relational preferences and center disabled people's sovereignty across individuals and contexts.

% --- acknowledgements ---
\begin{acks}
We would like to thank Cynthia Bennett, Cella Sum, Shivani Kapania, Ren Butler, Sanika Moharana, Xinru Tang, Julia Liu, Soyon Kim, Juno Bartsch, Frank Elavsky, and Ningjing Tang for their support through this project. We also are extremely grateful to our anonymous reviewers for their engaged feedback through the review process.
\end{acks}

% --- bibliography ---
\bibliographystyle{ACM-Reference-Format}
\bibliography{main}

\end{document}